\documentclass[review,12pt, fleqn]{elsarticle}

\usepackage{graphicx,amssymb}
\journal{International Journal of Mechanical Sciences}

\newcommand{\IR}{\mathbb{R}}

\begin{document}
\begin{frontmatter}
\title{Analytic description and explicit parametrization of the equilibrium shapes of elastic rings and tubes under uniform hydrostatic pressure}
\author[IMech]{P.A. Djondjorov\corref{cor}}
\ead{padjon@imbm.bas.bg}
\author[IMech]{V.M. Vassilev}
\ead{vasilvas@imbm.bas.bg}
\author[BioPhysics]{I.M. Mladenov}
\ead{mladenov@obzor.bio21.bas.bg}
\cortext[cor]{Corresponding author; tel. +359 2 979 64 78; fax +359 2 870 74 98}
\address[IMech]{Institute of Mechanics, Bulgarian Academy of Sciences \\ 
Acad. G. Bonchev St., Block 4, 1113 Sofia, Bulgaria}
\address[BioPhysics]{Institute of Biophysics, Bulgarian Academy of Sciences \\ 
Acad. G. Bonchev St., Block 21, 1113 Sofia, Bulgaria}

\begin{abstract}
The parametric equations of the plane curves determining the equilibrium shapes that a uniform inextensible elastic ring or tube could take subject to a uniform hydrostatic pressure are presented in an explicit analytic form. The determination of the equilibrium shape of such a structure corresponding to a given pressure is reduced to the solution of two transcendental equations. The shapes with points of contact and the corresponding (contact) pressures are determined by the solutions of three transcendental equations. The analytic results presented here confirm many of the previous numerical results on this subject but the results concerning the shapes with lines of contact reported up to now are revised.
\end{abstract}


\end{frontmatter}


\section{Introduction}

In the present paper, the problem for determination of the equilibrium shapes of a circular inextensible elastic ring subject to a uniformly distributed external force that acts normally to the ring in the ring plane is addressed. This problem is also referred to as the stability problem or buckling of the circular shape of the ring and the other equilibrium shapes are called buckled \cite{Carrier, TO, Chaskalovic}. It is also known (see, e.g., \cite{Flaherty, Antman, Pozrikidis}), that if a cylindrical elastic shell of circular cross section (i.e., a tube) is subject to a uniform external pressure, which is normal to its middle surface, then the typical cross section of the deformed tube takes the same shapes as the axis of a deformed elastic ring does provided that the latter is a simple curve (i.e., a curve without intersections).
Therefore, here the term ``ring" will be used to indicate both a proper ring and a tube.
It should be noted also that in the majority of the works in this field, the distributed force acting on a ring is called pressure as in the case of a shell. Following this tradition, we will use the same term in the present study remembering that pressure means force per unit length in the case of a ring and force per unit area in the case of a shell.  

Maurice L\'{e}vy \cite{Levy} was the first who stated and studied the problem under consideration and reduced the determination of the foregoing equilibrium shapes in polar coordinates to two elliptic integrals for the arc\-length and polar angle regarded as functions of the squared radial coordinate. He found also several remarkable properties of the equilibrium ring shapes and obtained that if the pressure $p$ is such that $p<(9/4)(D/\rho^3)$,
where $D$ and $\rho$ are the ring bending rigidity and radius of the undeformed shape, respectively,
then the ring possesses only the circular equilibrium shape.

Later on, Halphen \cite{Halphen} and Greenhill \cite{Greenhill} derived exact solutions to this problem in terms of Weierstrass elliptic functions on the ground of complicated analyses of the properties of the aforementioned elliptic integrals. Halphen (see \cite[p. 235]{Halphen}) found out that non-circular shapes with $n\geq2$ axes of symmetry are possible only for pressures greater than
$p_n = ( n^2 - 1) (D/ \rho^3)$.
Halphen \cite{Halphen} and Greenhill \cite{Greenhill} presented also several examples of non-circular equilibrium ring shapes. It should be noted, however, that the exact solutions reported in \cite{Halphen, Greenhill}, representing the polar angle as a function of the radius, appeared to be intractable and many researchers continued searching exact solutions \cite{Carrier, Zhang, Adams, Watanabe, DVM, VDM, JPhysA}, while others used various approximations \cite{TO, Flaherty, Pozrikidis, Wu} on the way to determine the equilibrium shapes of the ring.

Carrier \cite{Carrier} was the first who reconsidered the foregoing problem for the buckling of an elastic ring about half a century after the works by L\'{e}vy, Halphen and Greenhill. He expressed the curvature of the deformed ring in terms of Jacobi cosine function \cite{elliptic} involving several unknown parameters to be determined by a system of algebraic equations.
However, he succeeded to find approximate solutions to this system only for small deflections from the undeformed circular ring shape (see the exhaustive analysis provided recently by Adams \cite{Adams} who has criticized and developed Carrier's work \cite{Carrier}).

Tadjbakhsh and Odeh \cite{TO} studied the boundary-value problem describing the buckled shapes of the ring and the associated variational problem. They proved the existence of solutions to the boundary-value problem in the case of small deflections from the undeformed state and the existence of solutions of the associated variational problem (week solutions to the foregoing boundary-value problem) describing buckled shapes of an arbitrary deflection. 

Watanabe and Takagi \cite{Watanabe} thoroughly analyzed the variational problem for determination of the ring shapes stated by Tadjbakhsh and Odeh \cite{TO} and obtained analytic expressions for the curvature of the ring (in terms of Jacobi elliptic functions and Carrier's parameters \cite{Carrier}) and formulas for the slope angle at the points where the curvature has extrema. They also proved that non-circular shapes with $n\geq2$ axes of symmetry exist for pressures greater than
$p_n = ( n^2 - 1) (D/ \rho^3)$, thus extending Halphen's result,
and found out, moreover, that each such shape in unique.
This paper completes the branch of analysis of the considered problem that is based more or less on the approach suggested more than half a century ago by Carrier \cite{Carrier}. 

In the recent papers \cite{DVM, VDM, JPhysA}, the present authors have studied the differential equation for the curvature of the ring with the aim to achieve an analytic description of the cylindrical equilibrium shapes of lipid bilayer membranes. The determination of the analytic solutions of this equation reported in \cite{JPhysA} does not follow Carrier's approach \cite{Carrier}. Instead of this, the explicit formulas for the curvature of the buckled shapes are obtained in forms similar to those suggested by Zhang \cite{Zhang} for lipid bilayer membranes and by Fukumoto \cite{Fukumoto} in the context of fluid mechanics.
In \cite{JPhysA}, the parametric equations of the directrices of the considered cylindrical surfaces are expressed in an explicit analytic form, the necessary and sufficient conditions for such a surface to be closed are derived and several sufficient conditions for its directrix to be simple or self-intersecting are given.

The equilibrium shapes of closed planar elastic loops subject to the constraints of fixed length and enclosed area are studied also in the works by Arreaga {\it et al.} \cite{ACCG}, Capovilla {\it et al.} \cite{CCG} and Guven \cite{GV}.

Flaherty {\it et al.} \cite{Flaherty} presented a numerical determination of the equilibrium shapes of elastic rings or tubes subject to uniform external pressure. They suggested a scenario for the evolution of the equilibrium shape as the pressure increases, however, some stages of this scenario are not confirmed here.

Recently, Wu {\it et al.} \cite{Wu} presented approximate solutions of the considered problem by means of a special Fourier expansion of the curvature. They developed a linearized algebraic system for the unknown coefficients of this Fourier expansion and thus obtained an approximation of the ring shape.

The aim of the present paper is to provide an analytic description of the equilibrium states of an elastic ring or tube subject to a uniform hydrostatic pressure going as far as possible, to develop efficient computational procedures completing this analytic description and to reexamine some of the most important results obtained in this field.

The layout of the paper is as follows. The statement of the problem is given in Section 2.
A concise derivation of the most important L\'{e}vy's results is given in Section~3
where the equation for the ring curvature and the parametric equations of the equilibrium shapes are derived as well.
Using these results, in Section 4, the symmetry of the equilibrium shapes is justified. 
All periodic solutions to the equation for the ring curvature and the expressions for the corresponding slope angles are presented in explicit analytic form in Section 5.
Two systems of transcendental equations are derived in Sections 6 and 7 allowing to determine the equilibrium shapes corresponding to a given pressure and to calculate the values of the pressure at which the equilibrium shapes possess points of contact. The results concerning the equilibrium shapes with lines of contact obtained in \cite{Flaherty} are reexamined in Section~8.

\section{Differential equations for the equilibrium states}

Let us consider a ring made of a homogeneous isotropic linearly elastic material and assume that it is represented by its middle axis.
Suppose now that the ring is subject to a uniform external pressure $p$ acting along the normal vector to its stress-free configuration. The analysis carried out in the present work is based on the following three assumptions:
(\textit{i}) the ring axis is inextensible;
(\textit{ii}) the pressure $p$ preserves its magnitude and always acts as an external uniformly distributed force along the inward normal vector to the deformed ring axis, i.e., it is a uniform (simple) hydrostatic pressure;
(\textit{iii}) the deformed ring axis is a regular closed plane curve $\Gamma$ parametrized by its arc\-length $s$.

Next, let the curve $\Gamma$ be given by means of the coordinates $x(s)$, $y(s)$ of its position vector $\mathbf{r}(s)$ with respect to a certain rectangular Cartesian coordinate frame in the Euclidean plane, i.e.,
$
\mathbf{r}(s)=x(s)\,\mathbf{i}+y(s)\,\mathbf{j},
$
where $\mathbf{i}$ and $\mathbf{j}$ are the unit vectors along the coordinate axes $x$ and $y$, respectively. Consequently, 
the unit tangent vector $\mathbf{t}(s)$ and the unit inward normal vector $\mathbf{n}(s)$ to the curve $\Gamma$ are given as follows
\begin{equation}
\mathbf{t}(s)=x'(s)\,\mathbf{i}+y'(s)\,\mathbf{j}, \qquad
\mathbf{n}(s)=-y'(s)\,\mathbf{i}+x'(s)\,\mathbf{j}.
\label{tanorm}
\end{equation}
Here and throughout this paper primes denote derivatives with respect to the arclength $s$. 
Recall that the foregoing unit tangent and normal vectors are related to the curvature $\kappa(s)$ of the curve $\Gamma$ through the Frenet-Serret formulas \cite{Coxeter}
\begin{equation}
\mathbf{t}'(s) = \kappa(s) \mathbf{n}(s),  \qquad
\mathbf{n}'( s) = - \kappa(s)\mathbf{t}(s).
\label{FS}
\end{equation}

Finally, let $M(s)$, $N(s)$ and $Q(s)$ denote the bending moment and the components of the stress resultant force $\mathbf{F}(s)$ along the tangent and normal vectors to the curve $\Gamma$, i.e.,
\begin{equation}
\mathbf{F}(s)=N(s)\mathbf{t}(s)+Q(s)\mathbf{n}(s).
\label{SV}
\end{equation}
Then, the particular constitutive equation relating the moment $M(s)$ with the curvature $\kappa(s)$ and the form of the stress-free configuration of the ring together with the system of differential equations
\begin{equation}
\mathbf{F}'(s) = -p\, \mathbf{n}(s),
\label{eq11}
\end{equation}
\begin{equation}
M'(s) = -\mathbf{F}(s)\cdot \mathbf{n}(s),
\label{eq13}
\end{equation}
representing the local balances of the force and moment, respectively, in accordance with the assumption (\textit{ii}), 
and the closure conditions following from the assumption (\textit{iii}), which, without loss of generality, may be written in the form
\begin{equation}
\mathbf{r}(L)=\mathbf{r}(0), \qquad \mathbf{t}(L)=\mathbf{t}(0),
\label{BC}
\end{equation}
where $L$ is the length of the deformed ring, determine entirely the equilibrium state of the ring under consideration (see, e.g., \cite[Ch.~4)]{Antman}). Here and throughout this paper the dot stands for dot (scalar) product of two vectors. 
Let us remark also that using Eq. (\ref{SV}) and the Frenet-Serret formulas (\ref{FS}) one can represent the system of differential equations (\ref{eq11}), (\ref{eq13}) in the scalar form
\begin{equation}
N'(s) = Q(s)\kappa(s),
\label{Eqil1}
\end{equation}
\begin{equation}
Q'(s) = -N(s)\kappa(s)-p,
\label{Eqil2}
\end{equation}
\begin{equation}
M'(s) = -Q(s).
\label{Eqil3}
\end{equation}

\section{Parametric equations for the equilibrium shapes}
\label{3}

The aim of this Section is to prove the most important facts concerning the problem under consideration established by Maurice L\'{e}vy in his memoir \cite{Levy} and to derive the parametric equations of the equilibrium ring shapes. 

Using the expression for the normal vector, see Eqs. (\ref{tanorm}),
one can integrate the equilibrium condition (\ref{eq11}) and express the force vector in the form
\begin{equation}
\mathbf{F}(s)=py(s)\,\mathbf{i}-px(s)\,\mathbf{j},
\label{Fxy}
\end{equation}
omitting the constant of integration since it is always possible to eliminate it by choosing the origin of the coordinate frame at a certain privileged point which, actually, is the one that L\'{e}vy called ``centre of the elastic forces", cf. \cite[$1^{\circ }$ (p. 9)]{Levy}.
Eq. (\ref{Fxy}) implies that at each point of the deformed ring configuration the magnitude of the force vector
$F(s)=\left| \mathbf{F}(s)\right|$
is proportional to the magnitude of the position vector
$r(s)=\left| \mathbf{r}(s)\right|$,
the magnitude of the pressure $|p|$ being the coefficient of proportionality (cf. \cite[$2^{\circ }$ (p. 9)]{Levy}), i.e.,
\begin{equation}
F(s)= \left| p \right| r(s).
\label{Fpr}
\end{equation}
Next, taking the dot product of both sides of Eq. (\ref{Fxy}) with the normal vector one gets, bearing in mind the second one of equations (\ref{tanorm}), the relation
\[
\mathbf{F}(s)\cdot\mathbf{n}(s) =-p[x(s)x'(s)+y(s)y'(s)],
\]
which allows to integrate the balance of moment equation (\ref{eq13}) and to obtain
\begin{equation}
M(s)=\frac{p}{2}(r^2(s)+c),
\label{BMC}
\end{equation}
where $c$ is an arbitrary constant of integration, cf. \cite[$3^{\circ }$ (p. 9)]{Levy}. It is noteworthy that the relations (\ref{Fxy}) -- (\ref{BMC}) hold regardless of the particular material properties of the ring and the form of its stress-free configuration.

In terms of the slope angle $\varphi(s)$ one has the expressions
\begin{equation}
x'(s)=\cos \varphi(s), \qquad y'(s)=\sin \varphi(s),
\label{tns}
\end{equation}
\begin{equation}
\kappa(s)=\varphi'(s),
\label{ks}
\end{equation}
and using Eqs.~(\ref{tns}) can rewrite Eqs.~(\ref{tanorm}) in the form
\begin{equation}
\mathbf{t}(s)=\cos \varphi(s)\,\mathbf{i}+\sin \varphi(s)\,\mathbf{j},
\label{tanorm1}
\end{equation}
\begin{equation}
\mathbf{n}(s)=-\sin \varphi(s)\,\mathbf{i}+\cos \varphi(s)\,\mathbf{j}.
\label{tanorm2}
\end{equation}
Then, combining Eqs.~(\ref{SV}), (\ref{Fxy}), (\ref{tanorm1}) and (\ref{tanorm2}) one obtains the parametric equations of the deformed ring shape in the form
\begin{eqnarray}
&&x(s)= -\frac{1}{p}Q(s) \cos \varphi(s)-\frac{1}{p}N(s)\sin \varphi(s),
\nonumber \\
&&y(s)= -\frac{1}{p}Q(s) \sin \varphi(s)+\frac{1}{p}N(s)\cos \varphi(s).
\label{parametric}
\end{eqnarray}
Evidently, in view of Eq.~(\ref{tanorm1}), the second one of the closure conditions (\ref{BC}) implies that the rotation number of the deformed ring axis is $2m\pi$, where $m$ is an integer, i.e.,
\begin{equation}
\varphi(L)=\varphi(0)+2m\pi,
\label{BC1}
\end{equation}
whereas the first one of them transforms, on account of Eqs.~(\ref{BMC}), (\ref{parametric}) and (\ref{BC1}), into the obvious conditions for periodicity of the forces and moment 
\begin{equation}
N(L)=N(0), \qquad Q(L)=Q(0), \qquad M(L)=M(0).
\label{BC2}
\end{equation}
Again, neither the form of the parametric equations (\ref{parametric}) nor the forms of the boundary conditions (\ref{BC1}) and (\ref{BC2}) depend on
the particular material properties or the stress-free configuration of the ring.

Let us now assume that the constitutive equation of the ring is
\begin{equation} M(s)=D(\kappa(s) - \kappa^{\circ}), \label{eq14} \end{equation}
where $D$ is its bending rigidity and $\kappa^{\circ}=1/\rho$ is the curvature of its stress-free configuration, which is supposed to be a circle of radius $\rho$.
Using Eq.~(\ref{eq14}) and the Frenet-Serret formulas (\ref{FS}) we can immediately integrate the system of differential equations (\ref{Eqil1}) -- (\ref{Eqil3}) obtaining the following expressions for the tangent and normal components of the stress resultant force $\mathbf{F}(s)$
\begin{equation}
N(s)=-\frac{D}{2}( \kappa^{2}(s) -2\mu), \qquad Q(s)=-D \kappa'(s),
\label{NQ}
\end{equation}
and the single ordinary differential equation for the ring curvature
\begin{equation}
\kappa'' (s) +  \frac{1}{2}\kappa^3 (s) - \mu \kappa(s) - \sigma = 0,
\label{eq2}
\end{equation}
where $\sigma = p/D$ and $\mu$ is an arbitrary constant of integration.
On the other hand, combining equations (\ref{eq14}), (\ref{Fpr}) and (\ref{BMC}) we obtain the relation
\[
N^2(s)+Q^2(s)=2pD(\kappa(s)-\kappa^{\circ})-p^2c,
\]
which, in view of Eqs.~(\ref{NQ}), implies
\begin{equation}
\kappa ^{\prime }(s) ^{2} =  P (\kappa(s)),
\label{JPhysA}
\end{equation}
where $P ( \kappa )$ is a fourth-order polynomial of the curvature $\kappa$ of the form
\begin{equation}
P(\kappa) = - \frac{1}{4}\kappa ^{4} + \mu \kappa ^{2}+2 \sigma \kappa+\varepsilon
\label{polynomial}
\end{equation}
whose free term
$\varepsilon=-2\sigma \kappa^{\circ}-\sigma^2 c -\mu ^{2}$
incorporates all the constants of integrations introduced so far. Actually, Eq.~(\ref{JPhysA}) is a first integral of Eq.~(\ref{eq2})
(see \cite[Sec.~2]{JPhysA} for more details).
In this context, $\varepsilon$ is viewed as an arbitrary constant of integration.

Each sufficiently smooth real-valued solution $\kappa(s)$ of an equation of form (\ref{JPhysA}) corresponding to a certain triple of given values of the parameters $\mu$, $\varepsilon$ and $\sigma \neq 0$ generates, up to a rigid motion in the plane, a unique plane curve $\Gamma $ of curvature $\kappa(s)$.
The components of the position vector of this curve can by expressed in the form
\begin{eqnarray}
&&x(s) =\frac{1}{\sigma} \kappa'(s) \cos \varphi (s) + \frac{1}{2\sigma }( \kappa\, ^{2}(s)-2 \mu )
\sin \varphi (s),
\nonumber \\
&&y(s) =\frac{1}{\sigma} \kappa'(s) \sin \varphi (s) - \frac{1}{2\sigma} ( \kappa \,^{2}(s)-2 \mu )
\cos \varphi (s),
\label{parametric1}
\end{eqnarray}
obtained by substituting Eqs.~(\ref{NQ}) in the general formulas (\ref{parametric}).
However, the parametric equations (\ref{parametric1}) describe a shape that a ring of bending rigidity $D$ could take being subject to pressure $p=\sigma D$ if and only if the regarded solution $\kappa(s)$ of the respective equation of form (\ref{JPhysA}) is such that the closure conditions (\ref{BC1}) and (\ref{BC2}) hold for $L=2\pi\rho$; note that the latter equality follows from the assumption (\textit{i}).
If this is the case, then the respective solution $\kappa(s)$, its first derivative $\kappa'(s)$ and its indefinite integral
\begin{equation}
\varphi(s)=\int\kappa(s) \mathrm{d}s ,
\label{angle}
\end{equation}
cf. Eq.~(\ref{ks}), determine entirely the equilibrium state of the pressurized ring. Indeed, the shape of the ring is determined explicitly by Eqs.~(\ref{parametric1}) and the values of the moment and forces acting along the ring axes are given by the formulas (\ref{eq14}) and (\ref{NQ}), respectively.

It should be remarked that each such solution $\kappa(s)$ is necessarily a periodic function with period $L=2\pi\rho$, due to the condition (\ref{BC2}), and if $T$ is its least period, then $L = n T $, where $n$ is a positive integer. Since $\varphi (n T) =  n \varphi ( T)$, as follows by formula (\ref{angle}), the closure condition (\ref{BC1}) takes the form
\begin{equation}
\varphi(T) = \frac {1} {n} \varphi(0)+ \frac {2m \pi}{n}\,\cdot
\label{closure}
\end{equation}

\section{Symmetry of the ring shapes}

The symmetry of the ring shapes is discussed, usually without going into much detail, by almost all authors that contributed to the solution of the considered problem. A concise justification of this property is given below.

Let $\Gamma$ be a curve whose curvature $\kappa(s)$ is a periodic solution of Eq.~(\ref{JPhysA}) with least period $T$.
For each integer $i$ denote by $\Gamma^{(-)}_{it}$ and $\Gamma^{(+)}_{it}$ the parts of the curve $\Gamma$ corresponding to $s \in [\hat{T}_{i}-t ,\hat{T}_{i}]$ and $s \in [\hat{T}_{i}, \hat{T}_{i}+t]$, respectively, where $\hat{T}_i=i(T/2)$ and $t \in \IR$.
Since the relations
\begin{eqnarray}
 &&{\bf t}(\hat{T}_{i}) \cdot {\bf r}(\hat{T}_{i}+s) =
-{\bf t}(\hat{T}_{i}) \cdot {\bf r}( \hat{T}_{i}-s) \label{t} \\
 &&{\bf n}(\hat{T}_{i}) \cdot {\bf r}( \hat{T}_{i}+s) =
 {\bf n}(\hat{T}_{i}) \cdot {\bf r}( \hat{T}_{i}-s) 
\label{n}
\end{eqnarray}
hold for each couple of numbers $i \in \mathbb{N}$ and $s \in \IR$, the curves $\Gamma^{(-)}_{it}$ and $\Gamma^{(+)}_{it}$ are symmetric with respect to the axis directed by the normal vectors ${\bf n}(\hat{T}_{i})$. Fig.~1 illustrates the simplest case $i=1$, $t=\hat{T}_{1}$.
\begin{figure}[h]
\centering
\includegraphics[width=8cm]{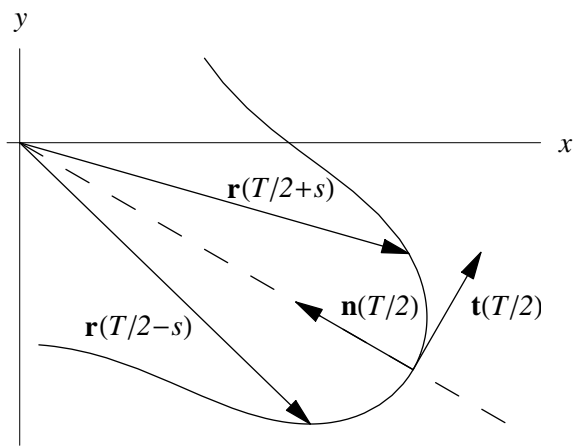}
\caption{Position vectors ${\bf r}(T/2\pm s)$, tangent ${\bf t}(T/2)$ and normal ${\bf n}(T/2)$ vectors to a curve; the dashed line represents the axis of symmetry.}
\end{figure}

It should be noted that relations (\ref{t}) and (\ref{n}) are easily verified using the parametric equations (\ref{parametric1}) and the formulas
\begin{equation}
\kappa (\hat{T}_{i}+s) = \kappa (\hat{T}_{i}-s), \qquad
\kappa' (\hat{T}_{i}+s) = - \kappa'(\hat{T}_{i}-s)
\label{symk}
\end{equation}
and
\begin{equation}
\varphi (\hat{T}_{i}+s)=\varphi (\hat{T}_{i})+\varphi (s)
\label{syma}
\end{equation}
that follow from $\kappa(-s)=\kappa(s)$ and the definition (\ref{angle}) of the slope angle $\varphi(s)$.

In other words, the curve $\Gamma$ can be thought of as generated by successive reflections of its part $\Gamma^{(-)}_{1t}$, $t=\hat{T}_{1}$, about the axes directed by the normal vectors ${\bf n}(\hat{T}_{i})$, $i \in \mathbb{N}$.
Apparently, if the curve $\Gamma^{(-)}_{nt} \cup \Gamma^{(+)}_{nt}$ get closed smoothly for some positive integer $n$, which may happened only for $t=\hat{T}_{n}$, then the closed curve $\Gamma=\Gamma^{(-)}_{nt} \cup \Gamma^{(+)}_{nt}$ has $n$ distinct axes of symmetry and is said to have $n$-fold symmetry or to be of $n$ mode.

\section{Analytic expressions for the ring curvatures and slope angles}

As it was mentioned in the Introduction, Eq.~(\ref{JPhysA}) was studied by the present authors with the aim to describe the cylindrical equilibrium shapes of fluid lipid bilayer membranes (see \cite{DVM, VDM, JPhysA}). Explicit analytic expressions for all periodic solutions of Eq.~(\ref{JPhysA}) and for the corresponding slope angles (\ref{angle}) are presented in \cite{JPhysA}. These results are outlined in this Section.

Depending on the values of the parameters $\sigma$, $\mu$ and $\varepsilon$, there exist two cases in which the polynomial $P(\kappa)$ attains positive values and hence Eq.~(\ref{JPhysA}) has real-valued solutions: (I)~the polynomial $P(\kappa)$ has two simple real roots $\alpha,\beta \in \IR$, $\alpha <\beta$, and a pair of complex conjugate roots $\gamma ,\delta \in \mathbb{C}$, $\delta =\bar{\gamma}$; (II)~the polynomial $P(\kappa)$ has four simple real roots  $\alpha<\beta<\gamma <\delta \in \IR$.
In the first case, the polynomial $P(\kappa)$ is nonnegative in the interval $\alpha \leq \kappa \leq \beta $, while in the second one, it is nonnegative in the intervals
$\alpha \leq \kappa \leq \beta $ and $\gamma \leq \kappa \leq \delta $.

It should be noted that the roots $\alpha ,\beta ,\gamma $ and
$\delta $ of the polynomial $P(\kappa)$ can be expressed explicitly
through its coefficients $\mu$, $\sigma$ and $\varepsilon$ and vice versa. Indeed, after some standard algebraic manipulations (see, e.g., \cite{Korn}), one can find the following expressions for the roots of the polynomial $P(\kappa)$
\[
-\sqrt{\omega } \pm \sqrt{2 \mu -
2 \sigma \sqrt{\frac{1}{\omega }}- \omega }, \qquad
\sqrt{\omega } \pm \sqrt{2 \mu +
2 \sigma \sqrt{\frac{1}{\omega }}- \omega },
\]
where
\begin{eqnarray*}
&& \omega =\frac{(2 \mu+\zeta )^{2}-2^{2}3\varepsilon}{6 \zeta}, \\
&& \zeta = \sqrt[3]{2^2 3 \left( 3^{2} \sigma ^{2}+\sqrt{\chi }\right) -2^3 \mu \left(\mu^{2}+ 3^{2} \varepsilon\right) }\,, \\[2mm]
&& \chi = 2^{2} 3 \varepsilon\left[ \left(\mu ^{2}+\varepsilon\right)
^{2}-3^{2} \mu \sigma ^{2}\right]
-  3 \sigma ^{2}\left( 2^2 \mu ^{3}-3^{3} \sigma ^{2}\right).
\end{eqnarray*}
Then, one can denote properly each of the above expressions for the roots in accordance with the notation introduced in the cases (I) and (II), respectively. Simultaneously, by Vieta's formulas one obtains
\begin{equation}
\alpha +\beta +\gamma +\delta = 0,
\label{RCond}
\end{equation}
due to the absence of a term with $\kappa^3$ in the polynomial $P(\kappa)$,
see Eq.~(\ref{polynomial}), and
\begin{eqnarray}
&&\mu = \frac{1}{4}\left(\alpha^2+\beta^2+\gamma^2+ \alpha\beta+\alpha \gamma+\beta \gamma\right),
\label{Coeffs1}\\
&&\sigma = -\frac{1}{8}\left( \alpha+ \beta\right)
\left( \alpha + \gamma \right) \left( \beta + \gamma \right), \label{Coeffs2} \\
&&\varepsilon = \frac{1}{4}\alpha\beta\gamma (\alpha + \beta\ + \gamma).
\nonumber
\end{eqnarray}

\subsection{Case (I)}
\label{4_1} 

Let the parameters $\mu$, $\sigma$ and $\varepsilon$ be such that the polynomial $P(\kappa)$ has roots as in case (I), namely, two of them are real ($\alpha<\beta$) and the other two constitute a complex conjugate pair which, in view of relation (\ref{RCond}), can be written in the form
\begin{equation}
\gamma = - \frac{\alpha+\beta}{2} + \mathrm{i} \eta, \qquad
\delta = - \frac{\alpha+\beta}{2} - \mathrm{i} \eta,
\label{gamma_delta}
\end{equation}
where $\eta$ is a nonnegative real number.
In this case, Eq.~(\ref{JPhysA}) has periodic solutions if $\eta \neq 0$ or $\eta = 0$ and $(3\alpha+\beta) (\alpha+3\beta) > 0$, see \cite[Theorem 1]{JPhysA}.

Let $\eta \neq 0$ and hence the roots of the polynomial $P(\kappa)$ are simple. Denote
\begin{equation}
\lambda_{1}=\frac{1}{4}\sqrt{AB}, \qquad
k_{1}=\sqrt{\frac{1}{2}-\frac{4\eta
^{2}+\left( 3\alpha +\beta \right) \left( \alpha +3\beta \right) }{2AB}},
\label {u1}
\end{equation}
where
\begin{equation}
A=\sqrt{4\eta^{2}+\left( 3\alpha +\beta \right) ^{2}},\qquad
B=\sqrt{4\eta^{2}+\left( \alpha+3\beta \right) ^{2}} . 
\label {lab}
\end{equation}
Evidently, $A>0$, $B>0$, $\lambda_1>0$ and $0 < k_1 < 1$. In this case, each solution of Eq.~(\ref{JPhysA}) can be expressed by the function
\begin{equation}
\kappa_1 \left( s\right) =\frac{\left( A\beta +B\alpha \right) - \left(
A\beta -B\alpha \right) \mathrm{cn}(\lambda_1 s,k_1)}{\left( A+B\right)
- \left( A-B\right) \mathrm{cn}(\lambda_1 s,k_1)},
\label {FK3}
\end{equation}
which takes real values for each $s \in \IR$ and is periodic with least period $T_1=(4/\lambda_1)\mathrm{K}(k_1)$ due to the periodicity of the Jacobi function $\mathrm{cn}(\lambda_1 s,k_1)$.
Here,  $\mathrm{K}(\cdot)$ denotes the complete elliptic integral of the first kind. The corresponding slope angle (\ref{angle}) can be written in the form
\begin{eqnarray}
\varphi_1 \left( s\right) &=&\frac{A\beta -B\alpha}{A-B}s \nonumber \\
&+& \frac{\alpha -\beta }{2\lambda_1 \sqrt{k_1 ^2+C}}\arctan \left( \sqrt{k_1 ^2+C}\,\frac{\mathrm{sn}\left(\lambda_1 s , k_1\right) }{\mathrm{dn}\left(\lambda_1 s ,k_1\right) }\right) \nonumber \\
&+& \frac{\left( A+B\right) \left( \alpha -\beta \right)
}{2\lambda_1 \left( A-B\right) }\Pi \left(-C,
\mathrm{am}( \lambda_1 s,k_1 ) ,k_1 \right),
\label{Ang33}
\end{eqnarray}
where $\mathrm{\Pi}(\cdot, \cdot, \cdot)$ denotes the incomplete elliptic integral of the third kind and
\begin{equation}
C=\frac{(A-B)^2}{4 A B}\, \cdot
\label{C}
\end{equation}

Now, let $\eta = 0$ and $(3\alpha+\beta) (\alpha+3\beta) > 0$. Then, the polynomial $P(\kappa)$ has one double and two simple real roots. The curvature and the slope angle (\ref{angle}) are expressed in terms of elementary functions as follows
\[
\kappa _{2}\left( s\right) =\frac{\left( A\beta +B\alpha \right) -\left(
A\beta -B\alpha \right) \mathrm{cos}(\lambda_1 s)}{\left( A+B\right) -\left(A-B\right) \mathrm{cos}(\lambda_1 s)},
\]
\[
\varphi_2 \left( s\right) =\frac{A\beta -B\alpha }{A-B}s
+\frac{8( \alpha -\beta) }{A-B} \arctan \left( \sqrt{\frac{A}{B}}
\tan \left( \frac{1}{2}\lambda_1 s\right) \right)  \cdot
\]
where $\lambda_1$, $A$ and $B$ remain defined by formulas (\ref {u1}) and (\ref {lab}), respectively.

\subsection{Case (II)}

Let the parameters $\mu$, $\sigma$ and $\varepsilon$ be such that the polynomial $P(\kappa)$ has roots as in case (II), that is $\alpha<\beta<\gamma<\delta \in \IR$. Denote 
\[
\lambda_2=\frac{1}{4}\sqrt{\left( \gamma -\alpha \right) \left( \delta
-\beta \right) }, \qquad
k_2=\sqrt \frac{\left( \beta -\alpha \right)
\left( \delta -\gamma \right) }{\left( \gamma -\alpha \right)
\left( \delta -\beta \right) } \cdot
\]
Since $\gamma -\alpha > \beta -\alpha > 0$ and $\delta
-\beta > \delta -\gamma >0$, it is seen that $\lambda_2 > 0$ and $0 < k_2 < 1$. In this case, each solution of Eq.~(\ref{JPhysA}) can be expressed by one of the following functions
\[
\kappa_3 (s)  =\delta -\frac{( \delta -\alpha )
( \delta -\beta ) }{( \delta -\beta )
+( \beta -\alpha ) \mathrm{sn}^{2}( \lambda_2 s,k_2)},
\]
\[
\kappa_4 (s)  =\beta +\frac{( \gamma -\beta )
( \delta -\beta) }{( \delta -\beta)
-( \delta -\gamma) \mathrm{sn}^{2}( \lambda_2 s,k_2)},
\]
see \cite[Theorem 2]{JPhysA}.
The functions $\kappa_3 (s)$ and $\kappa_4 (s)$ take real values for each $s \in \IR$ and are periodic with least period $T_2=(2/\lambda_2)\mathrm{K}(k_2)$ because of the periodicity of the function $\mathrm{sn}^{2}( \lambda_2 s,k_2)$. Their indefinite integrals (\ref{angle}) can be written as
\[
\varphi_3 (s) =\delta s-\frac{\delta -\alpha}{\lambda_2}\Pi
\left( \frac{\beta -\alpha }{\beta -\delta
},\mathrm{am}(\lambda_2 s,k_2),k_2 \right),
\]
\[
\varphi_4 (s) =\beta s-\frac{\beta -\gamma }{\lambda_2}\Pi
\left( \frac{ \delta -\gamma }{\delta -\beta },\mathrm{am}(\lambda_2s,k_2),k_2\right),
\]
respectively.

It should be remarked, that the indefinite integrals $\varphi_j (s)$, $j=1,\ldots,4$, of the foregoing solutions $\kappa_j (s)$ of Eq.~(\ref{JPhysA}) are chosen so that $\varphi_j (0)=0$.
Moreover, $\kappa_j (0)$ always coincides with a certain root of the polynomial $P(\kappa)$
(actually, $\kappa_j (0)=\alpha$ for $j=1,2,3$ and $\kappa_4 (0)=\gamma$)
and hence $\kappa_j ' (0) = 0$, according to Eq.~(\ref{JPhysA}).
It is important to note also that the functions $\kappa_j (s)$ are strictly increasing for $s \in [0, T/2]$ where $T$ is the respective least period. 

\section{Determination of the equilibrium shapes}

In the present study, our primary interest is in the determination of the equilibrium ring shapes that are curves without intersections, i.e., simple curves. It is established in \cite{JPhysA} that only the solutions of Eq.~(\ref{JPhysA}) that fall under the case (I) with $\eta \neq 0$ and $\alpha + \beta \neq 0$ may give rise to simple curves.
Therefore, hereafter we will restrict our analysis to
the regular closed curves $\Gamma_{n}$ of curvatures $\kappa(s)=\kappa_1 (s)$ given by formula (\ref{FK3}), which have $n$ axis of symmetry and meet all the necessary conditions for that to be simple. 

Thus, the closure condition (\ref{closure}) for such a curve $\Gamma_{n}$ reads
\begin{equation}
\varphi_1\left( T_1 \right) = \pm \frac {2 \pi} {n},
\label{closure2}
\end{equation}
as $\varphi_1(0)=0$ (see the remark at the end of Section~5), $n \geq 2$ due to the four vertex theorem (see, e.g., \cite{4VT}) and $m=\pm1$ since the rotation number of a simple regular closed curve must be $\pm 2 \pi$, see \cite{Hopf}.
Note, however, that there exist regular closed curve with rotation number $\pm 2 \pi$, which are not simple.

Then, substituting the expression $T_1=(4/ \lambda_1) K(k_1)$ for the least period of the solutions (\ref{FK3}) of Eq.~(\ref{JPhysA}) in the general formula (\ref{Ang33}) for the corresponding slope angle $\varphi_1(s)$ one can rewrite the closure condition (\ref{closure2}) in the form
\begin{equation}
\frac{( A+B) ( \alpha -\beta) }{2\lambda_1 ( A-B) }\Pi ( -C,k_1)
+\frac{A\beta -B\alpha}{\lambda_1(A-B)} \mathrm{K}(k_1) = \pm \frac {\pi} {2n}\, \cdot
\label{closure1}
\end{equation}

Finally, substituting the same expression for the period $T_1$ in
the relation $L=nT_1$ in order to take into account that the length of the ring $L$ is fixed and does not change upon deformation, see assumption (\textit{i}), one obtains
\begin{equation}
\frac{1}{\lambda_1}\mathrm{K}(k_1) = \frac{\pi}{2n} \, \cdot
\label{L}
\end{equation}
after setting for simplicity, without loss of generality, $L=2 \pi$, i.e., $\kappa^{\circ}=\rho=1$.

The left-hand sides of Eqs.~(\ref{closure1}) and (\ref{L}) are functions of the parameters $\alpha$, $\beta$ and $\eta$, see Eqs.~(\ref {u1}), (\ref {lab}) and (\ref{C}).
However, for the aims of the present study it is convenient to express the parameters $\alpha$ and $\beta$, using formulas (\ref{Coeffs2}) and (\ref{gamma_delta}), in terms of the positive parameters $\sigma$, $\eta$ and $q$ as follows
\begin{equation}
\alpha = \frac{4 \sigma}{\eta^2 + q^2} - q, \qquad \beta = \frac{4 \sigma}{\eta^2 + q^2} + q \cdot
\label{alphabeta}
\end{equation}

Thus, given an integer $n\geq2$ and a pressure $p$ by means of the parameter $\sigma$ (called hereafter simply ``pressure") the problem for the determination of the foregoing equilibrium shapes of the ring corresponding to this pressure is reduced to the computation of the solutions $\eta$ and $q$ of the transcendental equations (\ref{closure1}) and (\ref{L}).

It is important to notice that this problem has no nontrivial solution if $0<\sigma\leq \sigma_{bn}$ and has a unique nontrivial solution if $\sigma > \sigma_{bn}$, see \cite[Theorem~2]{Watanabe}. Here, $\sigma_{bn}=n^2-1$ is the so-called buckling pressure and by a trivial solution we mean the one, which corresponds to the ring shape that is a circle of radius $\rho=1$.

Here, the transcendental equations (\ref{closure1}) and (\ref{L}) corresponding to given integer $n\geq2$ and pressure $\sigma > \sigma_{bn}$ are solved numerically in two steps using \texttt{Mathematica}$^{\circledR}$.
First, the two curves in the $(\eta,q)$ plane defined by Eqs.~(\ref{closure1}) and (\ref{L}) are plotted using the routine \texttt{ContourPlot} in order to identify roughly the values of the coordinates of their intersection point. Then, these values are put as starting values in the routine \texttt{FindRoot}, which is employed to obtain the solutions $\eta$ and $q$ of the system of equations (\ref{closure1}), (\ref{L}) with a sufficient accuracy.
Once such a solution is determined, formulas (\ref {alphabeta}), (\ref{FK3}), (\ref{Ang33}), (\ref{C}) and the parametric equations (\ref{parametric1}) allow to depict the corresponding equilibrium ring shapes using the routine \texttt{ParametricPlot}.

Three examples of such shapes, which confirm the results presented in \cite{Flaherty} are given in Fig.~2.

\begin{figure}[h]
\centering
a \,
\includegraphics[width=1.2in,angle=90]{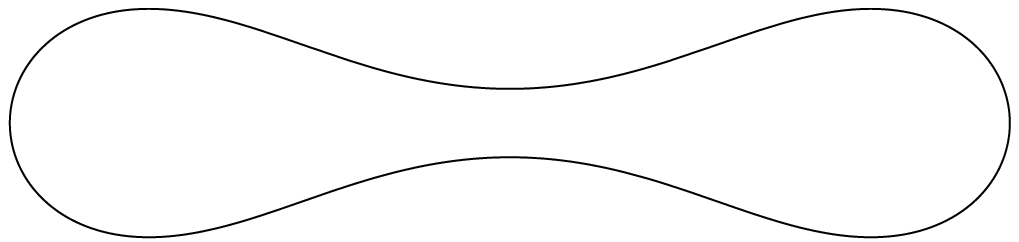} \quad
b
\includegraphics[height=1.2in]{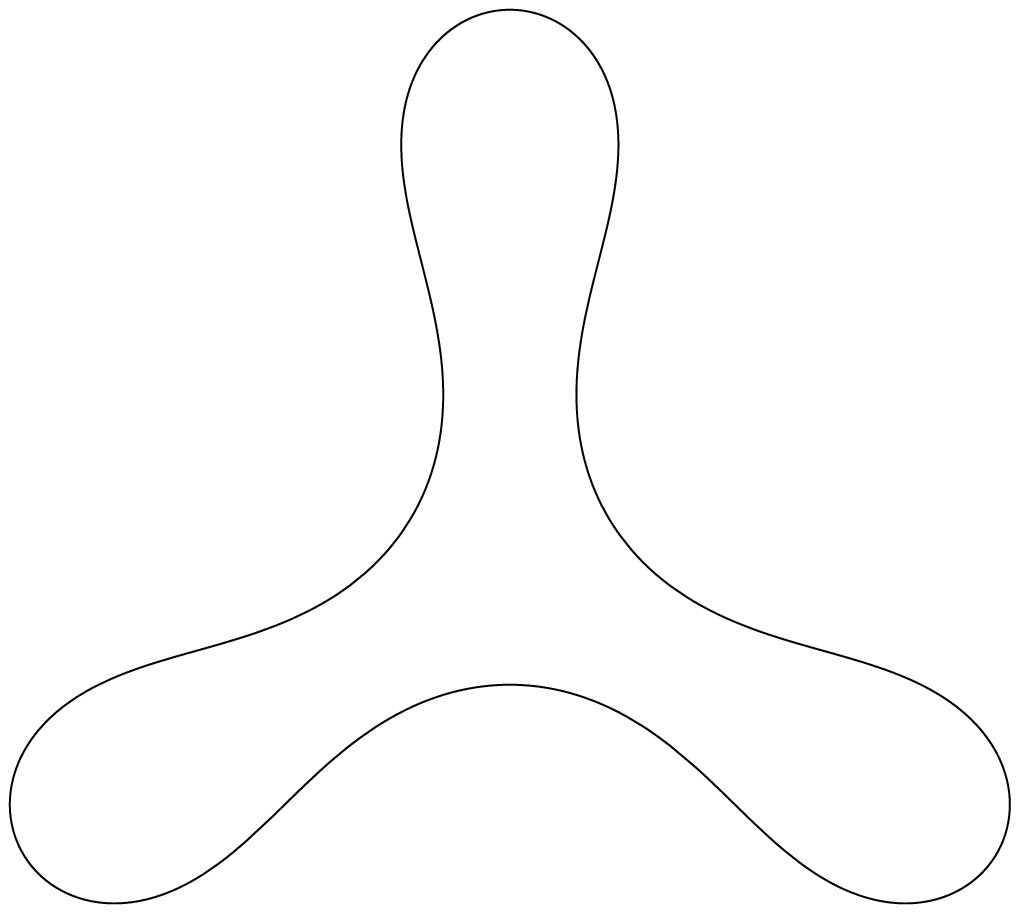}
\, \,
c
\includegraphics[height=1.2in]{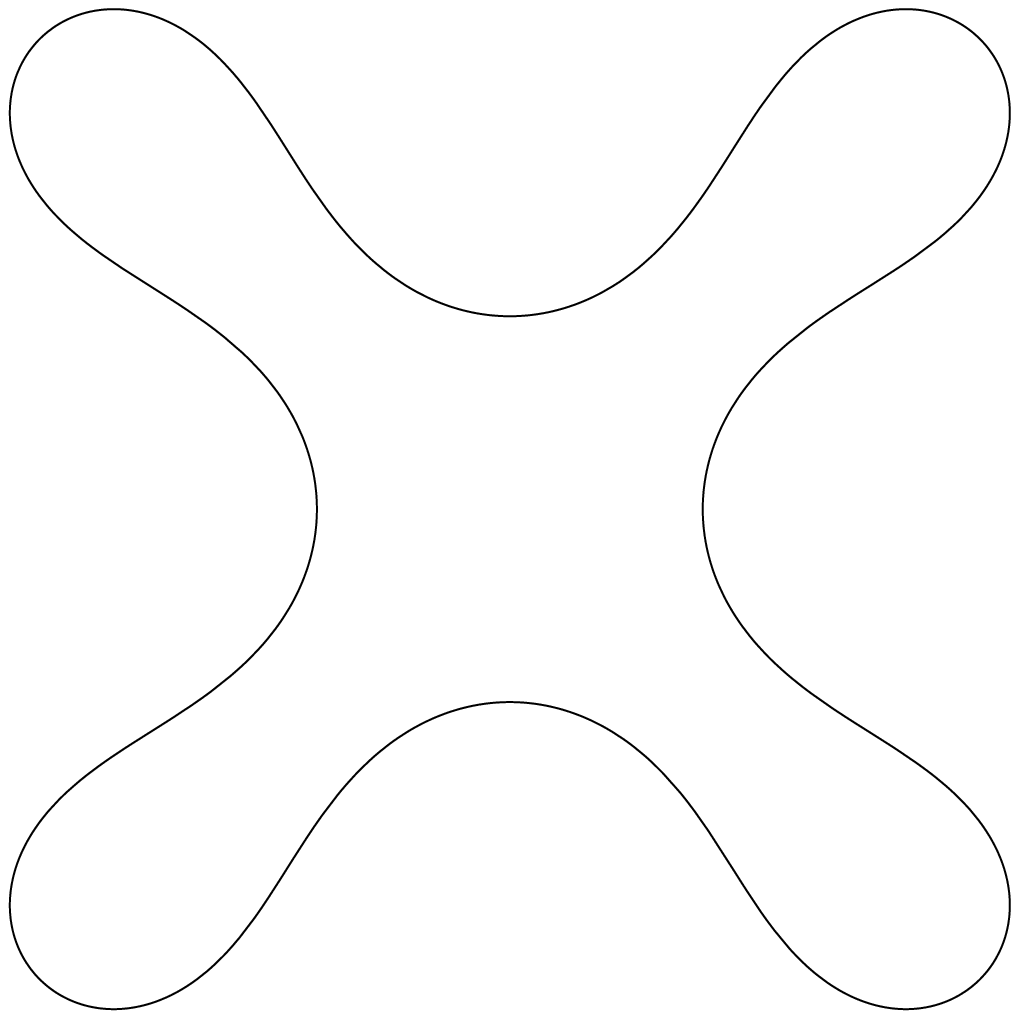}
\caption{Equilibrium ring shapes corresponding to: (a) $\sigma = 4.75$ (2-fold symmetry); (b) $\sigma = 16.25$ (3-fold symmetry); (c) $\sigma = 35.25$ (4-fold symmetry). }
\end{figure}

\section{Equilibrium shapes with points of contact}

In \cite{TO, Flaherty}, it is established that for each mode $n=2,3,4$ there is a value of the pressure $\sigma$, called contact pressure and denoted by $\sigma_{cn}$, at which some points of the respective buckled ring shape of $n$-fold symmetry come into contact.
In the aforementioned works, it is also observed that if the applied pressure $\sigma$ is such that $\sigma_{bn}<\sigma<\sigma_{cn}$, then the corresponding buckled shape of $n$ mode is simple.
It should be noted, that the values
for the contact pressures reported in \cite{TO, Flaherty} are obtained solving numerically a rather complicated nonlinear boundary-value problem.

In the present study, the determination of the non-circular equilibrium ring shapes with points of contact and the respective contact pressures is reexamined being reduced to the computation of the common solutions of the transcendental equations (\ref{closure1}), (\ref{L}) and one more algebraic, when $n=2$, or transcendental, when $n>2$, equation in the way to be described below.

Proceeding to the examination of this problem, let us first clarify, slightly extending the definition used in \cite{Flaherty}, that an $n$-mode
equilibrium ring shape $\Gamma_n$ is said to have a point of contact if it is not self-intersecting, but there is at least one couple of values $s_1$ and $s_2$ of the arc\-hlength $s$ such that $0<s_1 < s_2 < L$ and
\begin{equation}
{\bf r}(s_2) = {\bf r}(s_1), \qquad
{\bf t}(s_2) = \pm {\bf t}(s_1).
\label{contact}
\end{equation}
This means that at the point of contact ${\bf r}(s_2)={\bf r}(s_1)$ the curve $\Gamma_n$ is tangent to itself. Such a double point on a curve is called a cusp or tacnode, see~\cite{APD}.
The objective now is to reformulate the above conditions in a form suitable for the developing of an efficient procedure for computation of the contact pressures corresponding 
to the foregoing equilibrium ring shapes.

For that purpose, it is convenient to use the relations 
\begin{equation}
\kappa(s) =\frac{\sigma}{2}r^{2}(s)
-\frac{ \mu ^{2}+\varepsilon}{2\sigma }
\label{kr}
\end{equation}
and
\[
{\bf r}(s)\cdot {\bf t}(s)=\frac{1}{\sigma} \kappa'(s), \qquad
{\bf r}(s)\cdot {\bf n}(s)=- \frac{1}{2\sigma }( \kappa\, ^{2}(s)-2 \mu ),
\]
which follow from Eqs.~(\ref{JPhysA}) -- (\ref{parametric1}) and allow, taking into account Eqs.~(\ref{tanorm1}) and (\ref{tanorm2}), conditions (\ref{contact}) to be cast in the form
\begin{equation}
\kappa_{1}(s_2) =\kappa_{1}(s_1),\qquad
\kappa'_{1}(s_2)=\pm \kappa_{1}'(s_1),
\label{ds21}
\end{equation}
\begin{equation}
\kappa_{1} ^{2}(s_2)-2 \mu=\pm( \kappa_{1} ^{2}(s_1)-2 \mu ),
\label{kmu}
\end{equation}
and
\begin{equation}
\varphi_1(s_2)=\varphi_1(s_1)+2l\pi,
\label{m}
\end{equation}
if the sign in the second one of Eqs.~(\ref{contact}) is plus, or
\begin{equation}
\varphi_1(s_2)=\varphi_1(s_1)+(2l+1)\pi,
\label{l}
\end{equation}
if the foregoing sign is minus. Here, $l$ is an integer.

In the case when ${\bf t}(s_2) = {\bf t}(s_1)$, conditions (\ref{ds21}) imply $s_2=s_1+jT_1$, where $j$ is a positive integer and hence, according to Eqs.~(\ref{angle}) and (\ref{closure2}), we have
\begin{equation}
\varphi_1(s_2)=\varphi_1(s_1) \pm j\frac{2 \pi} {n} \cdot
\label{j}
\end{equation}
Now, combining Eqs.~(\ref{m}) and (\ref{j}) we obtain $j=\pm l n$, which means that $s_2=s_1\pm l n T_1=s_1\pm l L$ and therefore the assumptions ${\bf t}(s_2) = {\bf t}(s_1)$ and $0<s_1 < s_2 < L$ turn out to be  incompatible. That is to say that a non-circular equilibrium ring shape can not have cusps of this type.

In the remaining case ${\bf t}(s_2) = -{\bf t}(s_1)$, conditions (\ref{ds21}) and (\ref{kmu}) read
\begin{equation}
\kappa_{1}(s_2) =\kappa_{1}(s_1), \qquad
\kappa_{1}'(s_2)=- \kappa_{1}'(s_1),
\label{s211}
\end{equation}
\begin{equation}
\kappa_{1} ^{2}(s_2)-2 \mu =\kappa_{1} ^{2}(s_1)-2 \mu =0.
\label{kmu1}
\end{equation}
Now, taking into account the symmetry of the considered type of ring shapes, the invariance of Eq.~(\ref{eq2}) under the translation of its independent variable~$s$ and the particular properties of its solutions reflected by the formulas (\ref{symk}), without loss of generality one may assume that $s_1 \in [0,T_1/2]$ and can easily enumerate all the values~$s_2$ of the arclength for which conditions (\ref{s211}) hold and the part of the curve $\Gamma_n$ corresponding to $s \in [s_1,s_2]$ may have double points of the considered type only, namely: $s_2 = T_1 - s_1$ and $s_2 = nT_1-s_1$.
Hence, in view of Eqs.~(\ref{angle}) and (\ref{closure2}), condition (\ref{l}) takes the form
\begin{equation}
\varphi_1(s_1) = \pm \frac{\pi}{n}-\frac{\pi}{2}-l\pi,
\label{ll}
\end{equation}
if $s_2 = T_1 - s_1$, or
\begin{equation}
\varphi_1(s_1) = \pm \pi-\frac{\pi}{2}-l\pi,
\label{lll}
\end{equation}
if $s_2 = nT_1-s_1$.

Next, suppose that the curve $\Gamma_n$ is such that condition (\ref{kmu1}) holds for some $s \in [0,T_1/2]$. Actually, this means that in this interval there must be exactly two values of the arclength for which condition (\ref{kmu1}) is fulfilled, otherwise, by virtue of \cite[Theorem 3]{JPhysA}, the considered curve $\Gamma_n$ is self-intersecting that contradicts to the accepted definition for curves with points of contact.
Let us denote these values of the arclength by $s_1^-$ and $s_1^+$ and assume that
\begin{equation}
\kappa_{1}(s_1^-)=- \sqrt{2 \mu},\qquad
\kappa_{1}(s_1^+)= \sqrt{2 \mu}\,.
\label{kappa_mu_1_2}
\end{equation}
According to Eqs.~(\ref{kappa_mu_1_2}), there is some $s_0\in [0,T_1/2]$ for which $\kappa(s_0)=0$ and therefore, in the light of the above considerations,
$s_1^-< s_0 < s_1^+$ and $\kappa_1(0) \leq - \sqrt{2 \mu} < 0 < \sqrt{2 \mu} \leq \kappa_1(T_1/2)$ because the curvature $\kappa_1(s)$ is strictly increasing for $s \in [0,T_1/2]$ (see the remarks at the end of Section~5). Hence,
\begin{equation}
\kappa_{1} ^{2}(0)-2 \mu \geq 0, \qquad
\kappa_{1} ^{2}(T_1/2)-2 \mu \geq 0 \, .
\label{ineq}
\end{equation}
Taking into account that $\kappa_1'(0)=\kappa_1'(T_1/2)=0$ and $\varphi_1(0)=0$, the first of parametric equations~(\ref{parametric1}) implies
\begin{equation}
x(T_1/2)=\frac{1}{2\sigma}\left(\kappa_{1} ^{2}(T_1/2)-2 \mu \right) \sin\varphi_1(T_1/2).
\label{sign}
\end{equation}
The property (\ref{syma}) of the slope angle allows the closure condition (\ref{closure2}) to be recast in the form
\begin{equation}
\varphi_1(T_1/2)=\pm\frac{\pi}{n},
\label{closure3}
\end{equation}  
which, in view of the second of inequalities (\ref{ineq}) and the fact that $\sigma >0$, means that the sign of $x(T_1/2)$ coincides with that of the right hand side of Eq.~(\ref{closure3}). If the latter sign is minus, then $x(T_1/2)<0$ as implied by Eq.~(\ref{sign}) and the second of inequalities (\ref{ineq}). On the other hand, $x(0)=0$ and $x'(s)=\cos\varphi(s)$ meaning that $x(s)$ attains positive values in the interval $(0,T_1/2)$. Therefore, the respective curve $\Gamma_n$ is self-intersecting in this case because it intersects one of its symmetry axis -- the one directed by the normal vector ${\bf n}(0) = -{\bf j}$. In this way, we arrive at the conclusion that a contact point may occur only if the sign of the right hand side of closure condition Eq.~(\ref{closure3}) is plus.
\newpage

It is clear that the slope angle $\varphi_1(s)$ has a minimum at $s=s_0$ since $\varphi_1'(s_0)=\kappa_1(s_0)=0$ and $\varphi_1''(s_0)=\kappa_1'(s_0)>0$.
Actually, this is the only local extremum of this function in the interval $(0,T_1/2)$ since $s_0$ is the only point in the foregoing interval where the curvature is equal to zero.
Therefore, $\varphi_1(s)<\pi/n$ for $s \in (0,T_1/2)$ since $\varphi(0)=0$ and $\varphi(T_1/2)=\pi/n$. 

One can show also that $\varphi(s)>-\pi$ for $s \in (0,T_1/2)$. Indeed, if $\varphi (\hat{s})=-\pi$ for some $\hat{s} \in (0,T_1/2)$, then
\[
r'(\hat{s})=\frac{x( \hat{s}) \cos \varphi(\hat{s}) + y(\hat{s}) \sin \varphi (\hat{s})}{r(\hat{s})}=-\frac{x( \hat{s})}{r( \hat{s})}\leq 0
\]
since $x( s) \geq 0$ for each $s$ in the considered interval. However, this inequality contradicts the fact that $r(s)$ is strictly increasing for $s \in (0,T_1/2)$ as follows from Eq.~(\ref{kr}), and hence, $\varphi(s)>-\pi $ in this interval.

In view of the inequalities $-\pi<\varphi(s)\leq\pi/n $, conditions (\ref{ll}) and (\ref{lll}) take the form
\[
\varphi_1(s_1) = \frac{\pi}{n}-\frac{\pi}{2}, \qquad
\varphi_1(s_1) =-\frac{\pi}{2},
\]
respectively. Apparently, $\varphi_1'(s_1^-)<0$ and hence $\varphi_1(s)$ is decreasing in the neighbourhood of $s=s_1^-$, while $\varphi_1'(s_1^+)>0$ and hence $\varphi_1(s)$ is increasing at $s=s_1^-$. Therefore, contact points (if exist) should be such that
\begin{equation}
\varphi_1(s_1^-)=\frac{\pi}{n} - \frac{\pi}{2},
\label{cs1-}
\end{equation}
or
\begin{equation}
\varphi_1(s_1^+)= - \frac{\pi}{2},
\label{cs1+}
\end{equation}
otherwise the respective curve $\Gamma_n$ is self-intersecting.

Finally, taking the inverse 
\[
s=\frac{1}{\lambda_1} \mathrm{F}\left(\arccos \frac{A \beta + B \alpha - \kappa (A+B)}{A \beta - B \alpha - \kappa (A-B)} ,k_1 \right)
\]
of the function $\kappa=\kappa_{1}(s)$, which is readily achieved by Eq.~(\ref{FK3}) and well defined for each $s \in [0,T_1/2]$, one, bearing in mind that $s_1^-, s_1^+ \in [0,T_1/2]$ and Eqs.~(\ref{kappa_mu_1_2}), obtains
\begin{equation}
s_1^-=\frac{1}{\lambda_1} \mathrm{F}\left(\arccos \frac{A \beta + B \alpha + \sqrt{2 \mu} (A+B)}{A \beta - B \alpha + \sqrt{2 \mu} (A-B)} ,k_1 \right),
\label{s1-}
\end{equation}
\begin{equation}
s_1^+=\frac{1}{\lambda_1} \mathrm{F}\left(\arccos \frac{A \beta + B \alpha - \sqrt{2 \mu} (A+B)}{A \beta - B \alpha - \sqrt{2 \mu} (A-B)} ,k_1 \right),
\label{s1+}
\end{equation}
where $\mathrm{F}(\cdot, \cdot)$ is the incomplete elliptic integral of the first kind.

Thus, two triples of transcendental equations (\ref{closure1}), (\ref{L}) and (\ref{cs1-}) or (\ref{cs1+}) arise for the determination of the $n$-mode equilibrium ring shapes with points of contact. In both cases, the respective transcendental equations involve as unknowns only the four parameters $\sigma$, $\eta$, $q$ and $n$ since the archlengths $s_1^-$ and $s_1^+$, which may correspond to points of contact are determined explicitly in terms of these parameters by formulas (\ref{s1-}) and (\ref{s1+}), and the same holds true for the left hand sides of Eqs.~(\ref{cs1-}) and (\ref{cs1+}) in view of the general expression (\ref{Ang33}) for the slope angle. Of course, one should remember that the sign of the right hand side of Eq.~(\ref{closure1}) is plus in this context.

To summarize: given an integer $n \geq 2$, each solution of any one of the aforementioned two triples of transcendental equations gives the value of the contact pressure $\sigma_{cn}$ and the values of the parameters $\eta$ and $q$ determining in this way an equilibrium ring shape of $n$-fold symmetry with points of contact.
\newpage

Solving numerically the foregoing two systems of transcendental equations using the routine \texttt{FindRoot} in \texttt{Mathematica}$^{\circledR}$ we have found that the system consisting of Eqs.~(\ref{closure1}), (\ref{L}) and (\ref{cs1+}) does not have solutions for $2 \leq n \leq 15$, but the system of equations (\ref{closure1}), (\ref{L}) and (\ref{cs1-}) has a unique solution for each such mode $n$. Our conjecture is that this happens for all modes.

The obtained values for the respective contact pressures $\sigma_{cn}$ are presented in Table 1.
The equilibrium ring shapes with points of contact corresponding to the contact pressures $\sigma_{c2}$, $\sigma_{c3}$ and $\sigma_{c4}$ are depicted in Fig.~3.
  
\begin{center}
Table 1. Contact pressures.\phantom{Table 1. Contact pressures pressures......}
\medskip

\begin{tabular}{ccccccccccc}
\hline
$n $ & $2$ & $3$ & $4$ & $5$ & $6$ & $7$ & $8$ \\ \hline
$\sigma_{cn}$ & $5.247$ & $21.650$ & $51.844$ & $97.834$ & $161.077$ & $242.682$ & $343.517$\\ \hline
$n $ & $9$ & $10$ & $11$ & $12$ & $13$ & $14$ & $15$ \\ \hline
$\sigma_{cn}$ & $464.276$ & $605.522$ & $767.719$ & $951.253$ & $1156.450$ & $1383.580$ & $1632.890$ \\
\hline
\end{tabular}
\end{center}
\medskip \medskip

It is worth noting that in the special case $n=2$ the transcendental equation (\ref{cs1-}) may be replaced by the algebraic relation  \begin{equation}
\sigma=\frac{\left(\eta^2 + q^2\right)^2}{16q} \cdot
\label{sigma_c}
\end{equation}
Indeed, in this case, expression (\ref{cs1-}) simplifies to $\varphi_1(s_1^-) = 0$ meaning that $s_1^- = 0$. In fact, there exist two values of the archlength within the interval $[0,T_1/2]$ in which $\varphi_1(s) = 0$ but the other one is necessarily greater than $s_0$ and therefore it is disregarded. Then, the first of Eqs.~(\ref{kappa_mu_1_2}) reads
$ \alpha^2-2 \mu=0 $.
Substituting here expressions (\ref {alphabeta})$_1$ and (\ref{Coeffs1}) for the root $\alpha$ and the parameter $\mu$, respectively, and accounting for Eqs.~(\ref{gamma_delta}) and (\ref {alphabeta})$_2$ one obtains Eq.~(\ref{sigma_c}).

\begin{figure}[h]
\centering
a \,
\includegraphics[width=1.1in,angle=90]{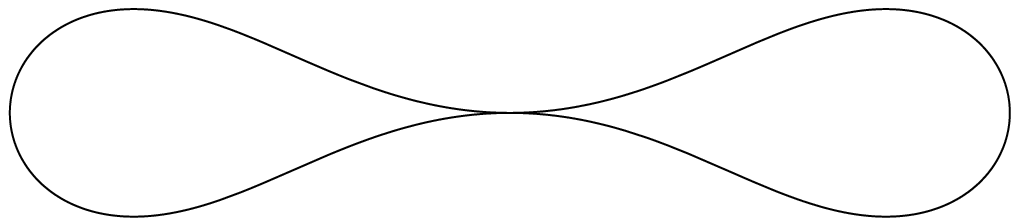}
\quad
b
\includegraphics[height=1.1in]{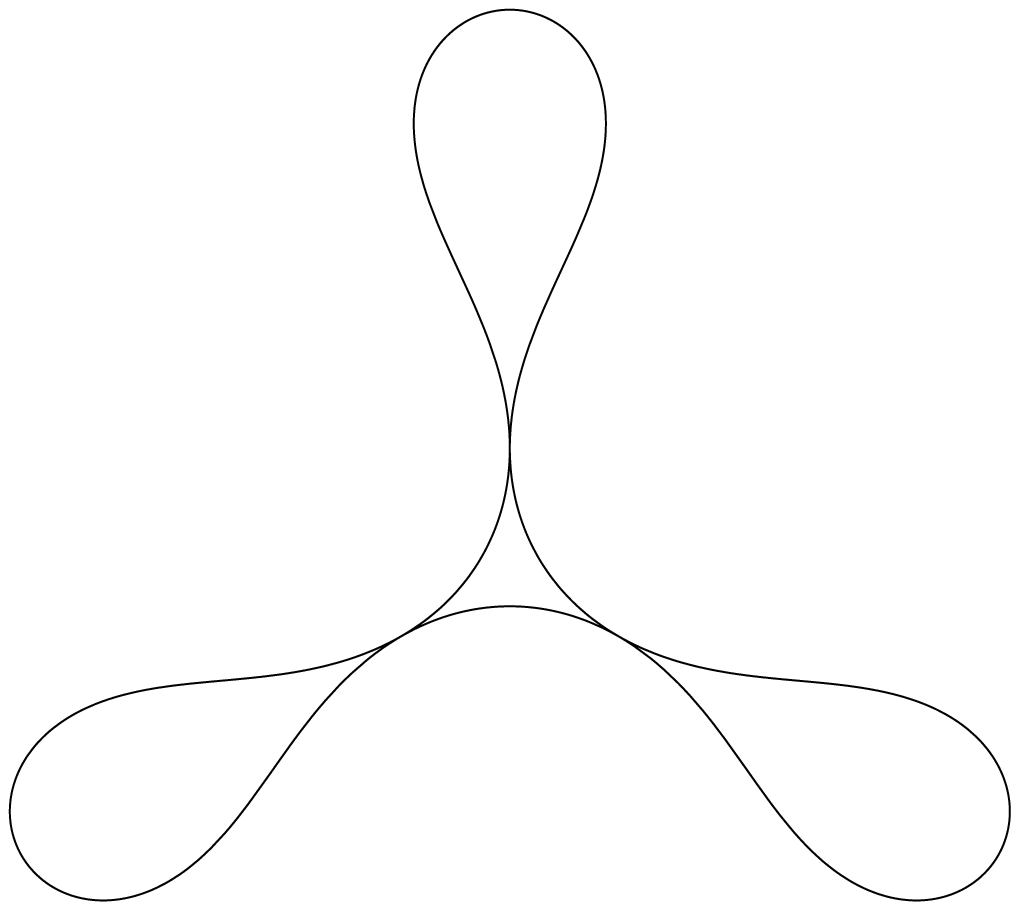}
\, \,
c
\includegraphics[height=1.1in]{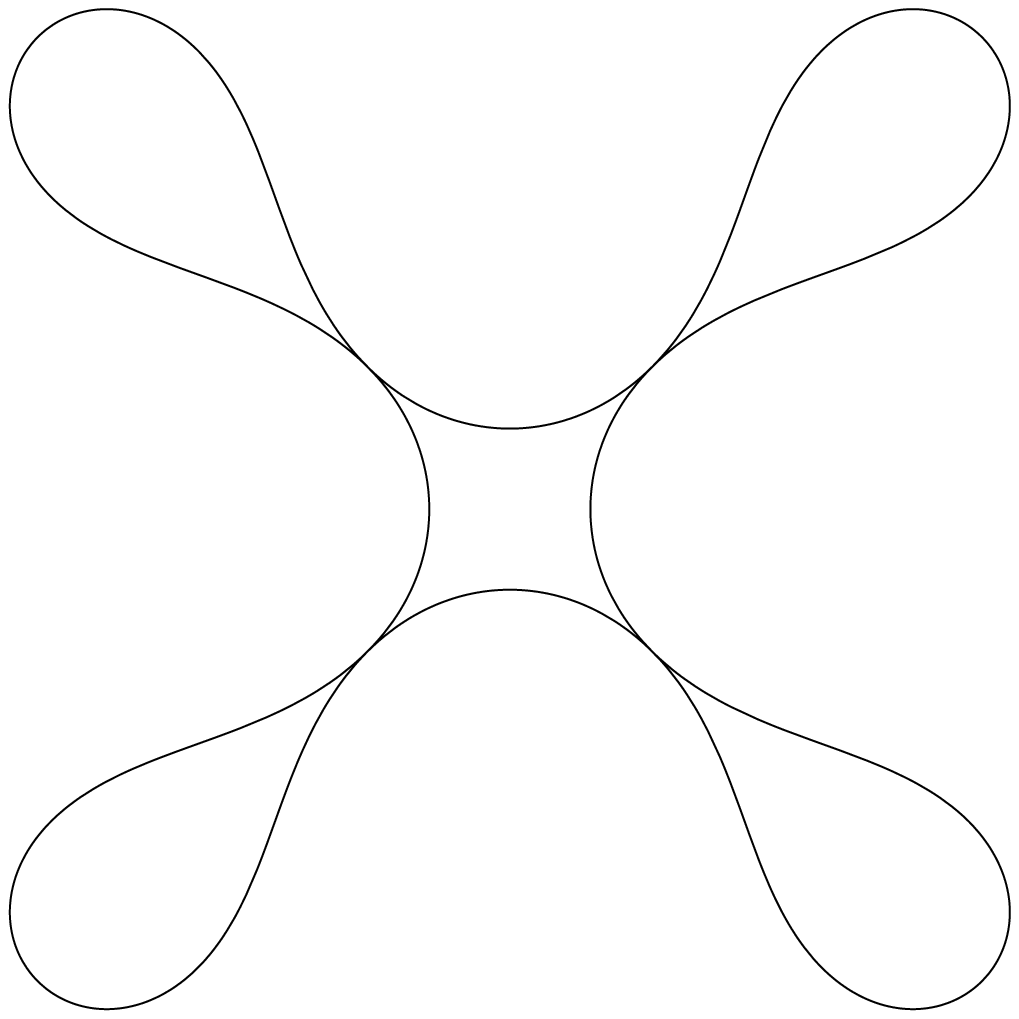}
\caption{Ring shapes with points of contact: (a) $\sigma = 5.247$;
(b) $\sigma = 21.65$; (c) $\sigma = 51.844$. }
\end{figure}

The values of contact pressures $\sigma_{cn}$ obtained for $n=2,3,4$ confirm exactly the results presented in \cite[formulas (2.11)]{Flaherty},
see also Fig.~3,
but the latter are interpreted therein as the lowest values of the pressures at which an isolated point of contact occurs. In \cite{Flaherty}, it is claimed that for each mode $n$ beyond the contact pressure $\sigma_{cn}$ there exists a continuous range of pressures, from $\sigma_{cn}$ up to a certain pressure denoted by $\sigma_{0n}$, such that for each $\sigma_{cn} \leq \sigma \leq \sigma_{0n}$ the respective ring shape exhibits contacts at isolated points only.
The pressure $\sigma_{0n}$ is set in \cite{Flaherty} to be the one for which the curvature at the corresponding contact point is zero.
If so, however, then in view of Eqs.~$(66)$ $\mu=0$ and $s_1^-=s_1^+=s_0$. Therefore, the equation $\kappa(s) - 2 \mu = 0$ has exactly one solution for $s \in [0,T_1/2]$ and hence, according to \cite[Theorem 3]{JPhysA}, the corresponding ring shape is self-intersecting.
Thus, in this respect the results of Flaherty {\it et al.} \cite{Flaherty} turn out to be inaccurate in spite of the fact that they are widely accepted and even confirmed numerically by other authors (see, e.g., \cite{BP}).

Actually, for all modes $n$ in the range $2 \leq n \leq 15$ our computations based on the procedure described in Section~6 show that if the applied pressure $\sigma$ is such that $\sigma_{bn}<\sigma<\sigma_{cn}$, then the corresponding buckled shape of $n$ mode is simple, while for $\sigma>\sigma_{cn}$ this shape always has points of self-intersection.
Our conjecture is that this behaviour is inherent to all modes.

For $n=2,3,4,$ Flaherty {\it et al.} \cite{Flaherty} affirm that for each pressure $\sigma$ such that $5.247 \leq \sigma \leq 10.34$, $21.65 \leq \sigma \leq 81.81$ or $51.84 \leq \sigma \leq 207.2$, respectively, the corresponding equilibrium ring shapes have only isolated points of contact without being self-intersecting. Our results presented in Figs.~4, 5 and 6 show that this is not the case. Let us recall that for each $\sigma > \sigma_{bn}$ the corresponding equilibrium ring shape of $n$-fold symmetry is unique, see \cite[Theorem~2]{Watanabe}.
\begin{figure}[h]
\centering
a \,
\includegraphics[width=1.2in,angle=90]{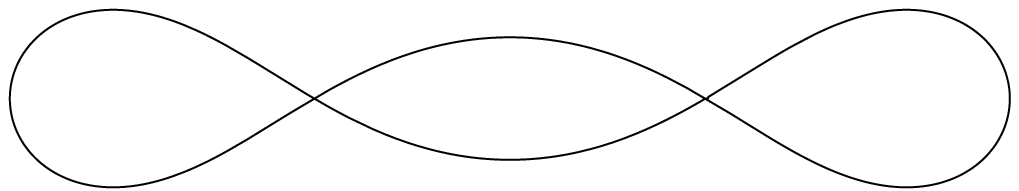} \quad
b \,
\includegraphics[width=1.2in,angle=90]{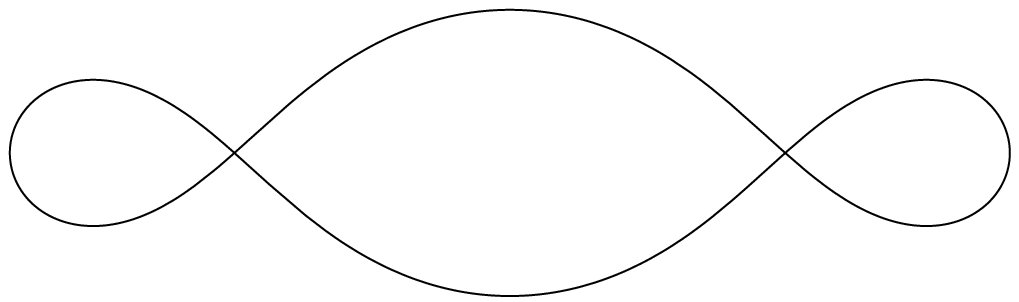} \quad
c \,
\includegraphics[width=1.2in,angle=90]{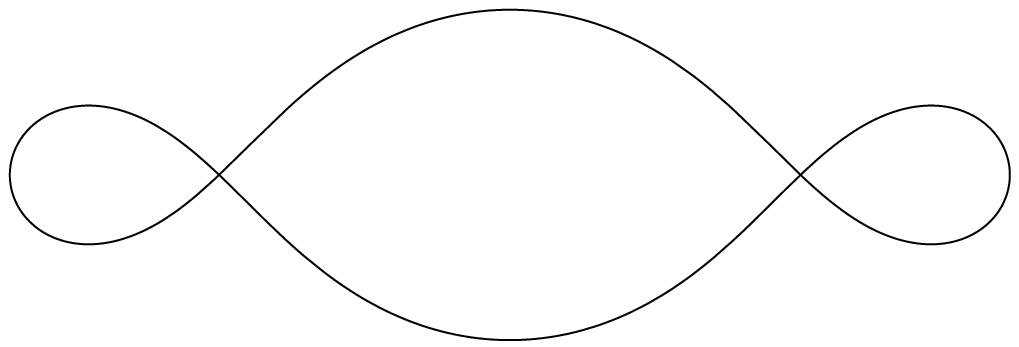}
\caption{Ring shapes corresponding to: (a) $\sigma = 6.48$;
(b) $\sigma = 9.24$; (c) $\sigma = 10.34$. }
\end{figure}~
\begin{figure}[h]
\centering
a \,
\includegraphics[width=1.2in]{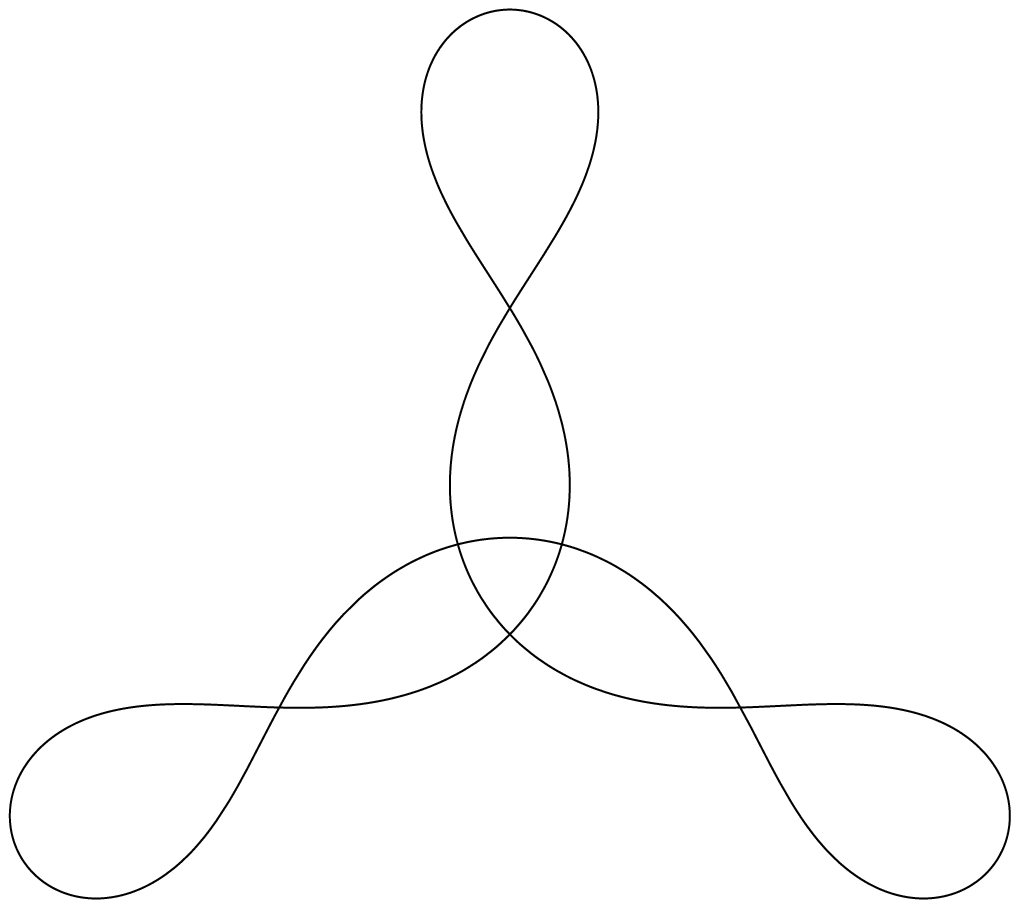} \quad
b \,
\includegraphics[width=1.2in]{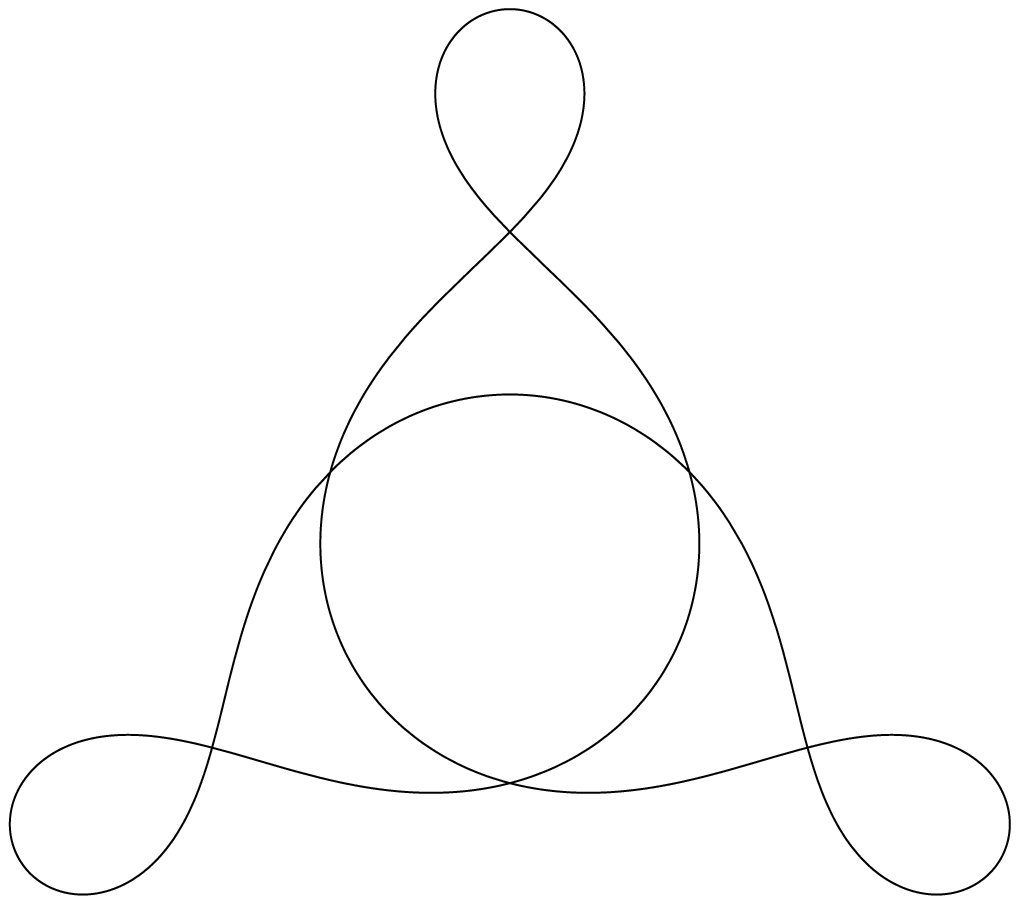} \quad
c \,
\includegraphics[width=1.2in]{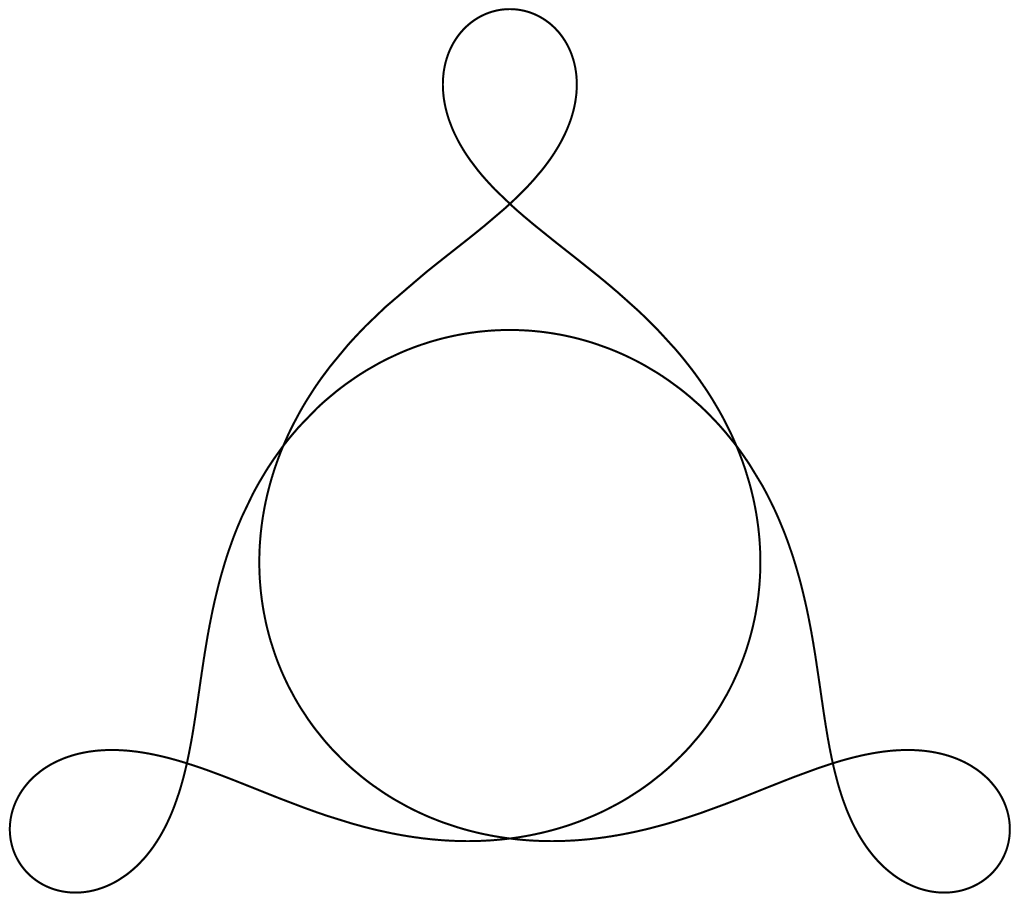}
\caption{Ring shapes corresponding to: (a) $\sigma = 28.56$;
(b) $\sigma = 56.09$; (c) $\sigma = 81.81$. }
\end{figure}~
\begin{figure}[h]
\centering
a \,
\includegraphics[width=1.2in]{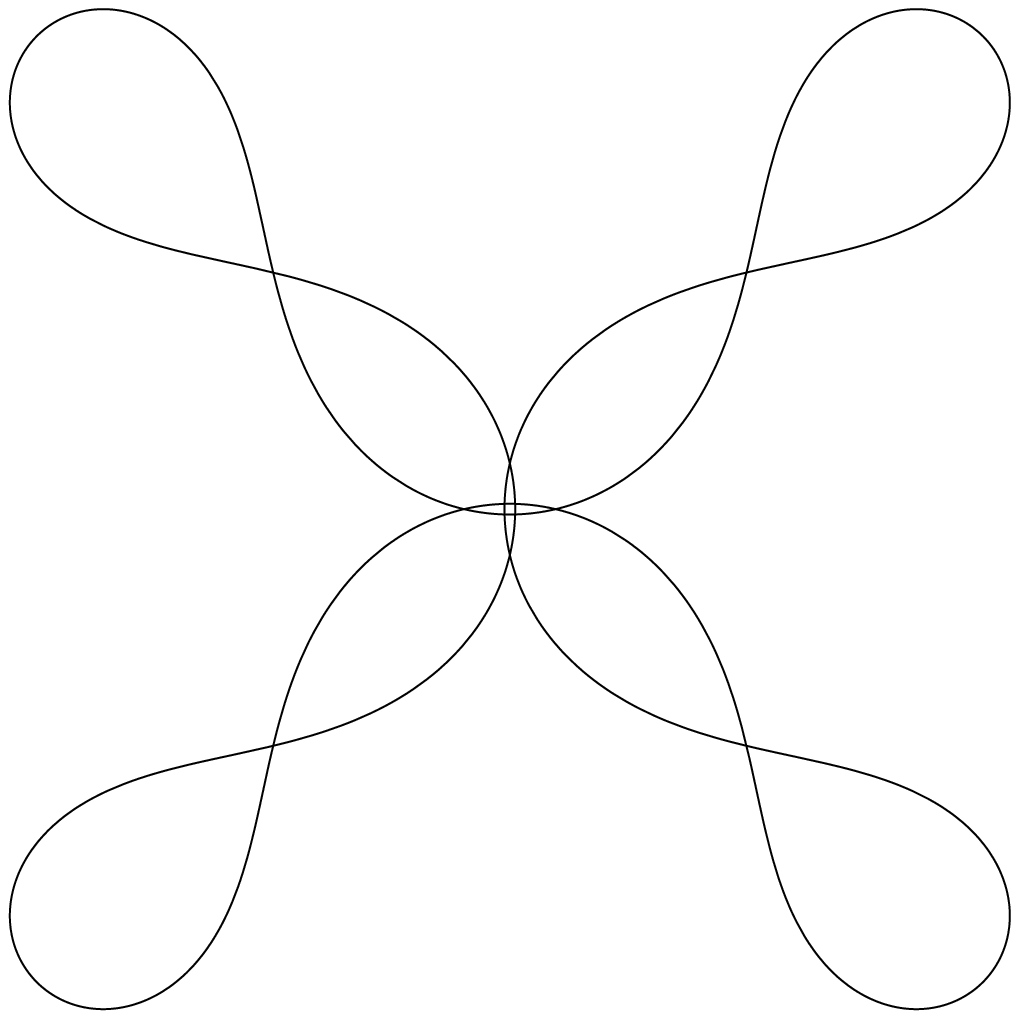} \quad
b \,
\includegraphics[width=1.2in]{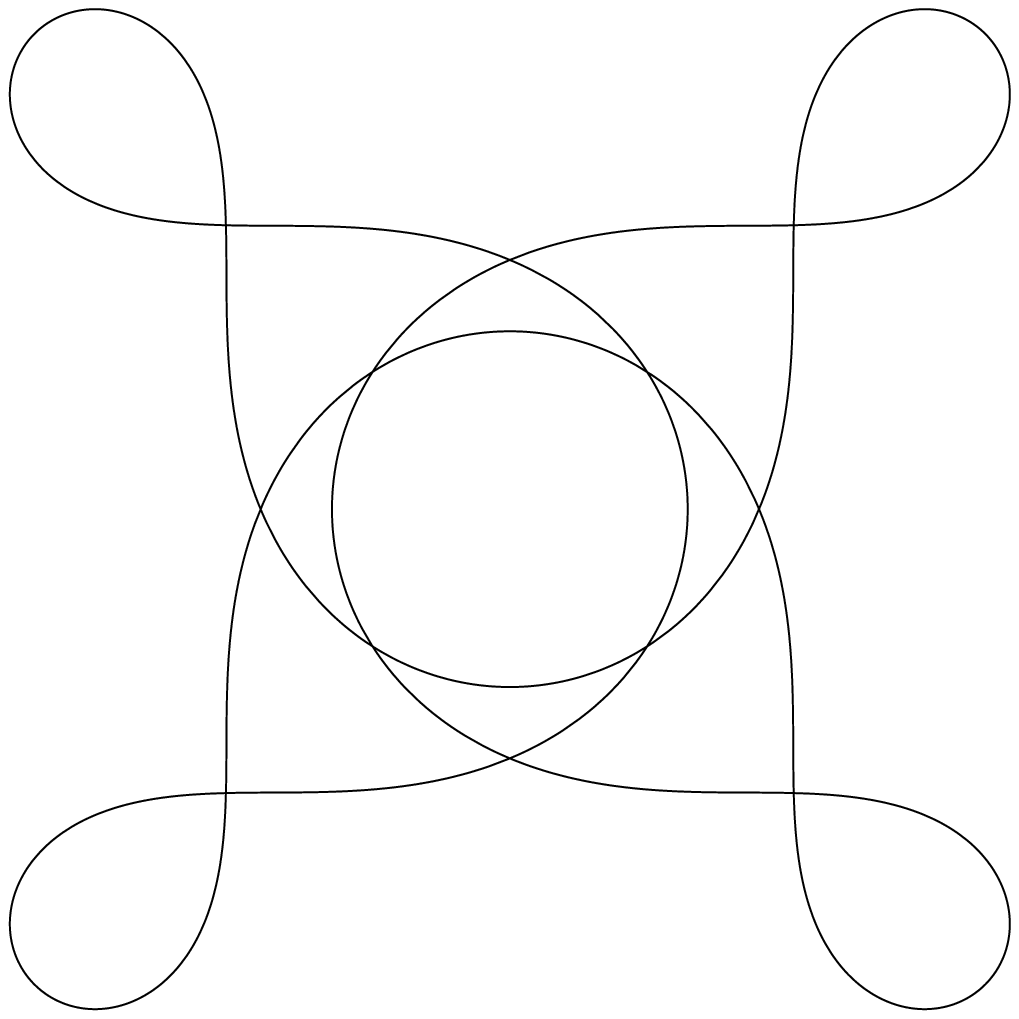} \quad
c \,
\includegraphics[width=1.2in]{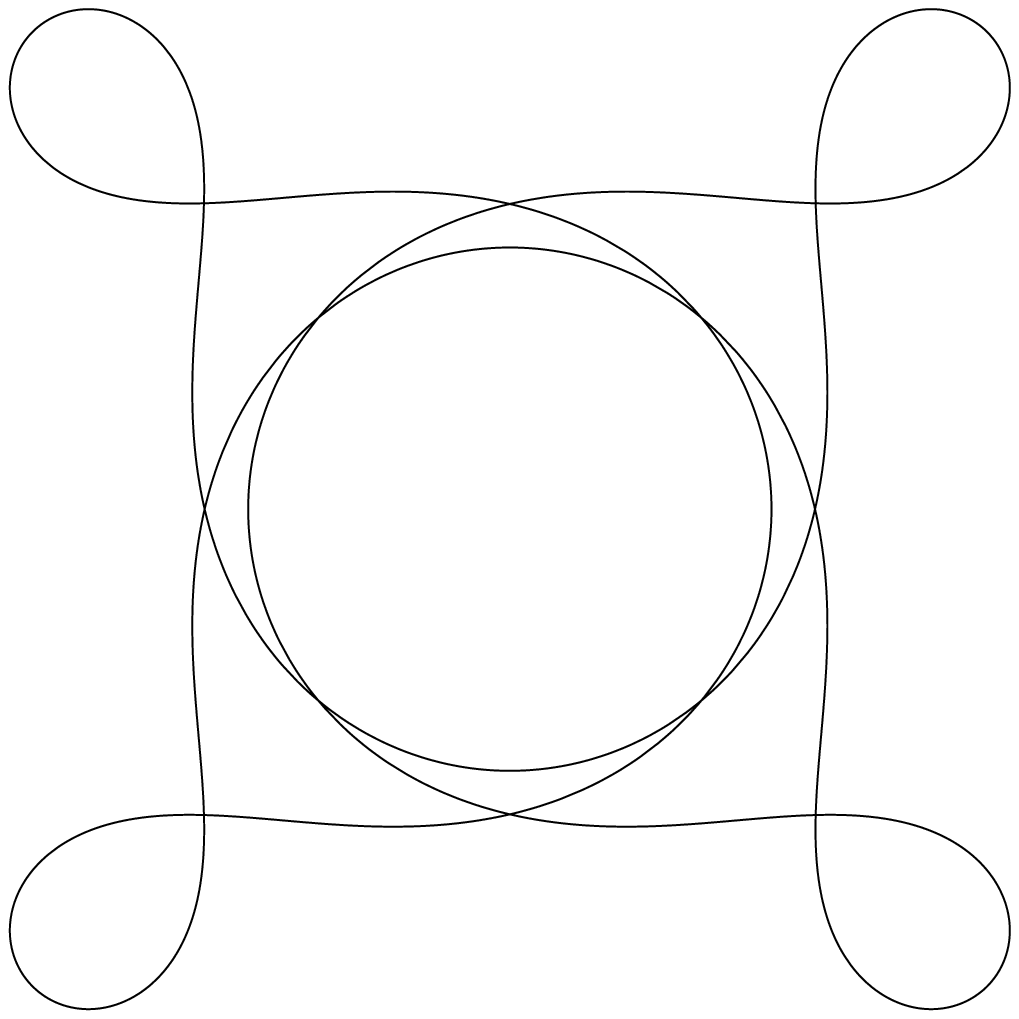}
\caption{Ring shapes corresponding to: (a) $\sigma = 70.7$;
(b) $\sigma = 140$; (c) $\sigma = 207.2$. }
\end{figure}

\section{Equilibrium shapes with lines (areas) of contact}

Apparently, for rings of finite thickness self-intersecting shapes are not possible because they are not planar, nevertheless for a very thin ring such a shape may be considered as a good approximation of its equilibrium state. 
A tube evidently can not take a self-intersecting shape, but there is a good reason to expect that tubes subject to sufficiently high pressure posses equilibrium shapes with areas of contact (lines of contact of their cross sections).

Flaherty {\it et al.} \cite{Flaherty} suggest similarity transformations to be used for the determination of such shapes, 
but realize this idea in a very complicated way.
Moreover, the construction developed in \cite{Flaherty} for the said purpose makes use of the curves $\Gamma_{0n}$ corresponding to the pressures $\sigma_{0n}$, which are wrongly regarded as curves with isolated points of contact as it was noted and discussed above.

Below, an alternative approach is presented for constructing equilibrium ring (tube) shapes with lines (areas) of contact based on the same ``similarity" idea that actually arises out of the following property of Eqs.~(\ref{eq2}) and (\ref{JPhysA}).

Under the transformation $(s,\kappa )\longmapsto (s/\lambda ,\lambda \kappa )$, where $\lambda $ is an
arbitrary real number, each equation of form (\ref{eq2}) corresponding to
certain constants $\mu $ and $\sigma $ transforms into an equation of the
same form but with new coefficients: $\mu
\longmapsto \lambda ^{2} \mu$, $\sigma \longmapsto \lambda ^{3} \sigma$. 
The same holds true for equation (\ref{JPhysA}) if $\varepsilon \longmapsto \lambda ^{4} \varepsilon$ in addition.
In other words, equations (\ref{eq2}) and (\ref{JPhysA}) are invariant with respect to the similarity transformation $\Lambda:(s,\kappa;\mu,\sigma,\varepsilon )\longmapsto (s/\lambda ,\lambda \kappa;\lambda ^{2}\mu,\lambda ^{3} \sigma,\lambda ^{4} \varepsilon)$. Consequently, the parametric equations (\ref{parametric1}) imply that the shapes whose parameters are related by such a transformation $\Lambda$ are similar, the respective scaling factor being $1/\lambda$. Accordingly, if a closed curve $\Gamma$ is scaled in this way, then its length $L$ and area $A$ change to $L/\lambda$ and $A/\lambda^2$, respectively.

Thus, given $n \geq 2$, let the curve $\Gamma_{cn}$ of length $L_{cn} = 2 \pi$ be the equilibrium shape with points of contact corresponding to the contact pressure $\sigma_{cn}$ and let $\hat{\Gamma}$ be the shape (of the same length) with lines of contact corresponding to a pressure $\hat{\sigma}>\sigma_{cn}$. The curve $\hat{\Gamma}$ is constructed in two steps. First, scaling the curve $\Gamma_{cn}$ with a factor $(\hat{\sigma} / \sigma_{cn})^{1/3}$ one obtains another curve $\hat{\Gamma}_{cn}$ which has the same number of contact points because it is similar to the curve $\Gamma_{cn}$ but corresponds to the pressure $\hat{\sigma}$ and its length is $\hat{L}_{cn}=2 \pi(\sigma_{cn}/\hat{\sigma})^{1/3} < L_{cn}$. Then, the curve $\hat{\Gamma}$ is obtained by substituting each point of contact of the curve $\hat{\Gamma}_{cn}$ by a line segment of length $2 \pi (1 -(\sigma_{cn}/\hat{\sigma})^{1/3})/n$ along the respective symmetry axis of the curve $\hat{\Gamma}_{cn}$  so as its total length to become $2\pi$.

Examples of shapes with lines of contact are presented in Figs.~7 and 8.
\begin{figure}[h]
\centering
a \includegraphics[width=1.2in,angle=90]{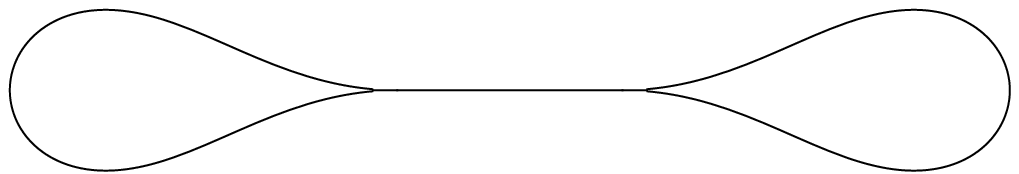} \quad
b \includegraphics[height=1.2in]{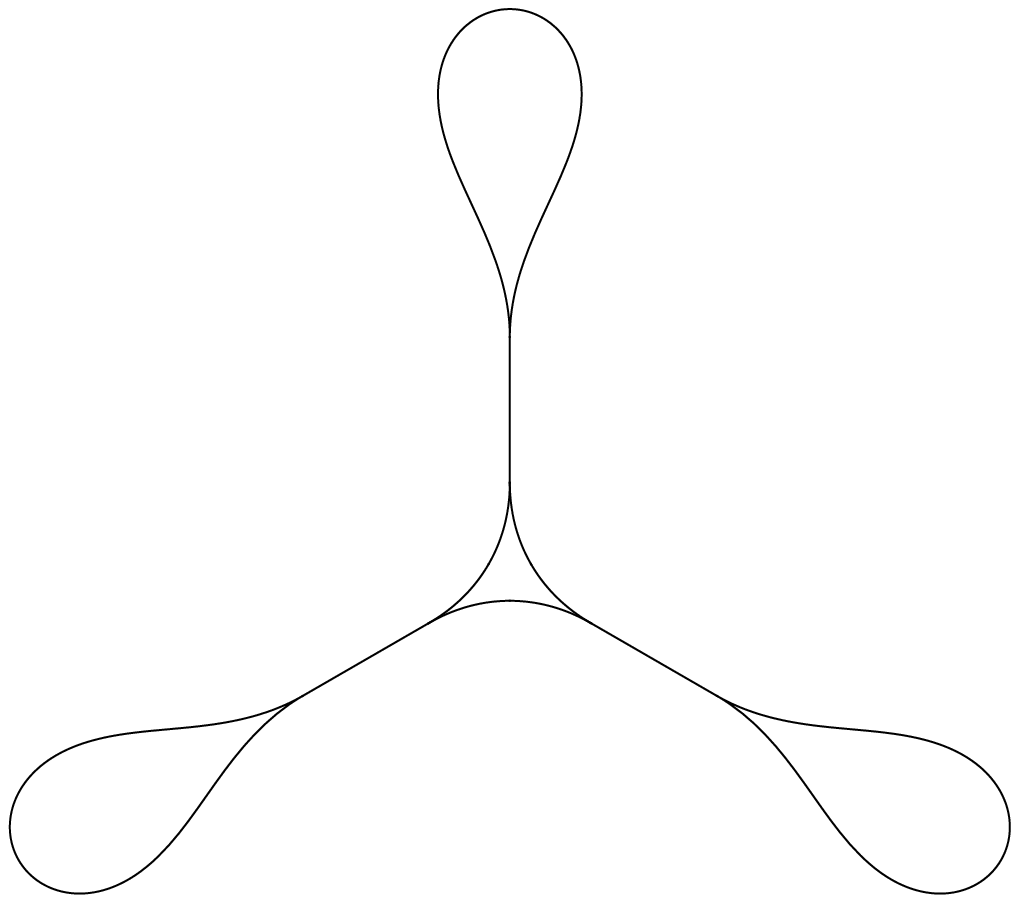} \,
c \includegraphics[height=1.2in]{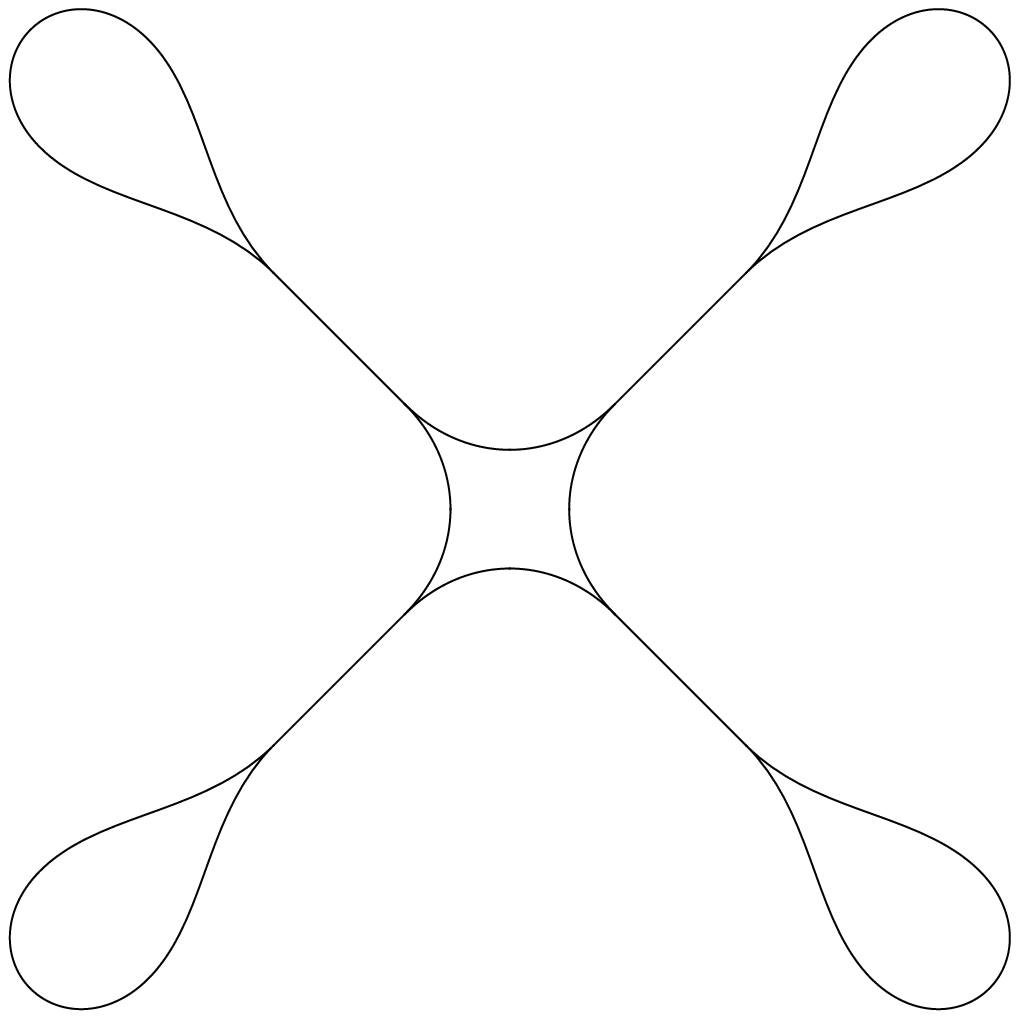}
\caption{Shapes with lines of contact corresponding to:
(a) $\sigma = 10.34$ (2-fold symmetry); (b) $\sigma = 81.81$ (3-fold symmetry); (c) $\sigma = 207.2$ (4-fold symmetry).}
\end{figure}
\vspace{-0.7cm}
\begin{figure}[h]
\centering
a \includegraphics[width=1.2in,angle=90]{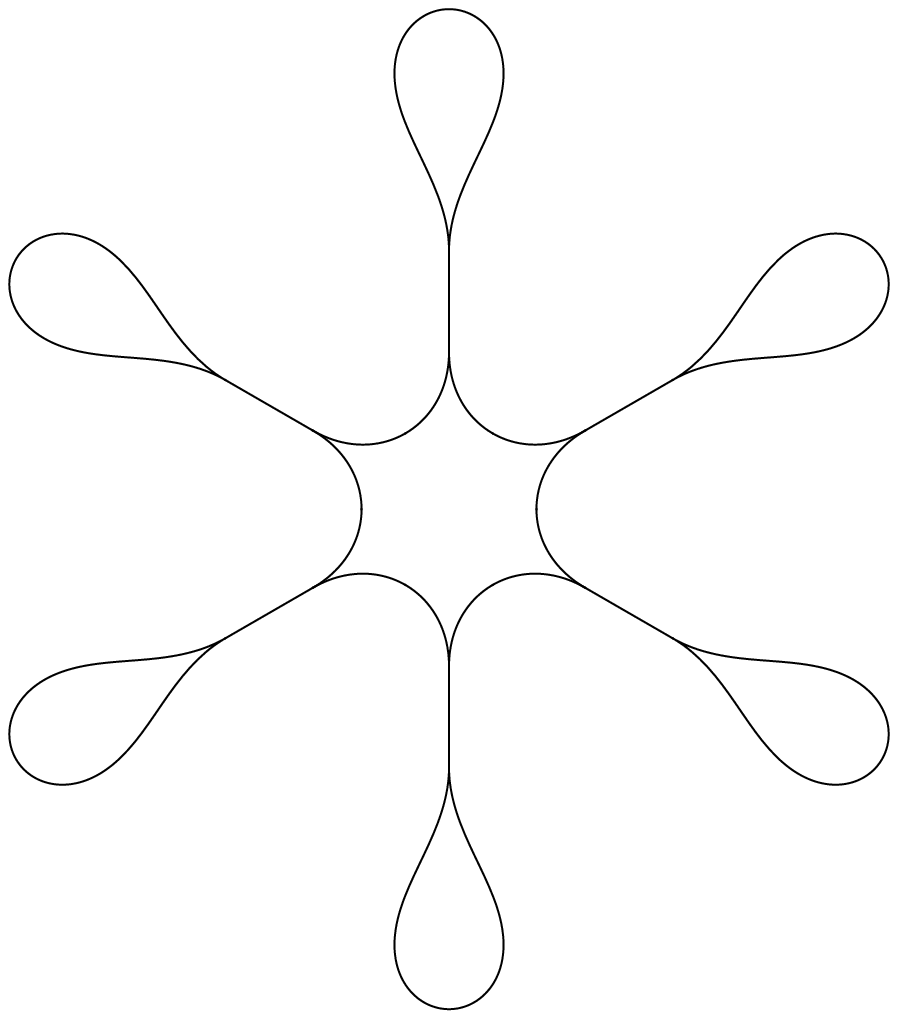} \quad
b \includegraphics[height=1.2in]{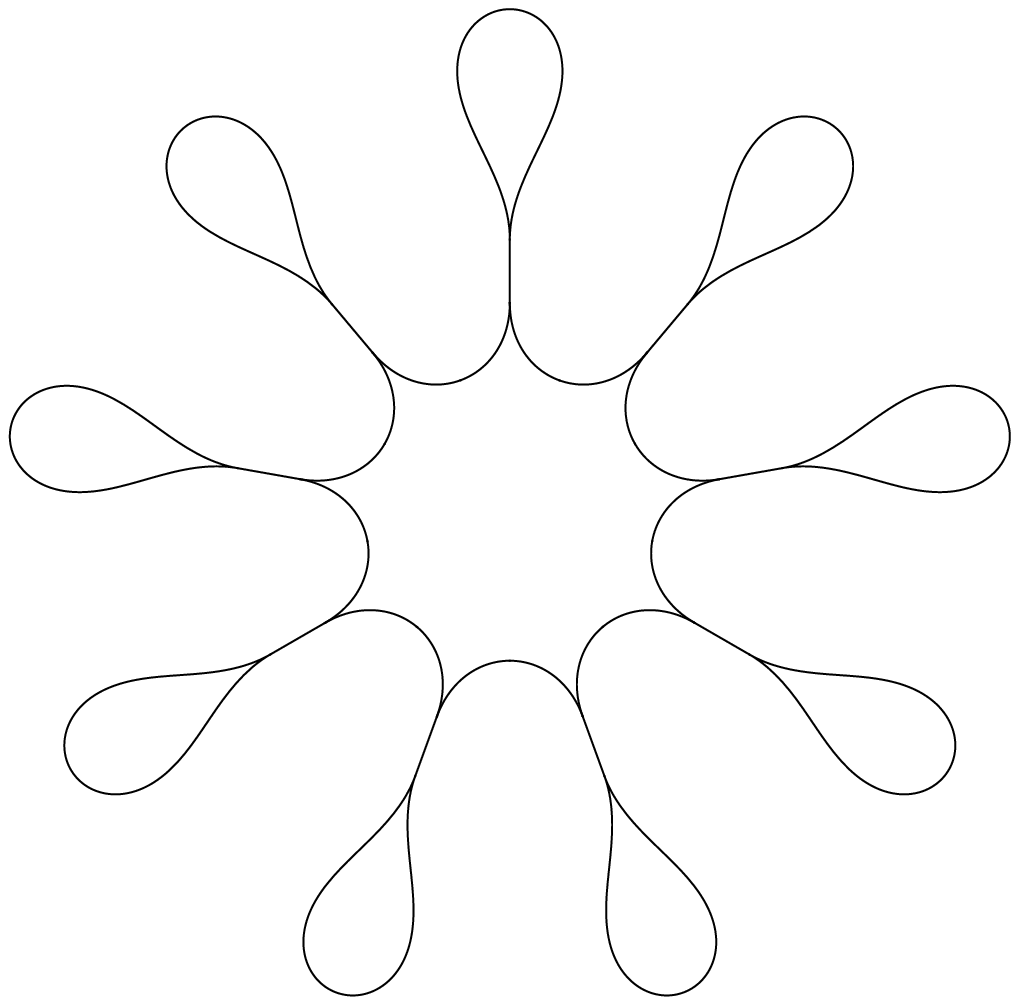} \,
c \includegraphics[height=1.2in]{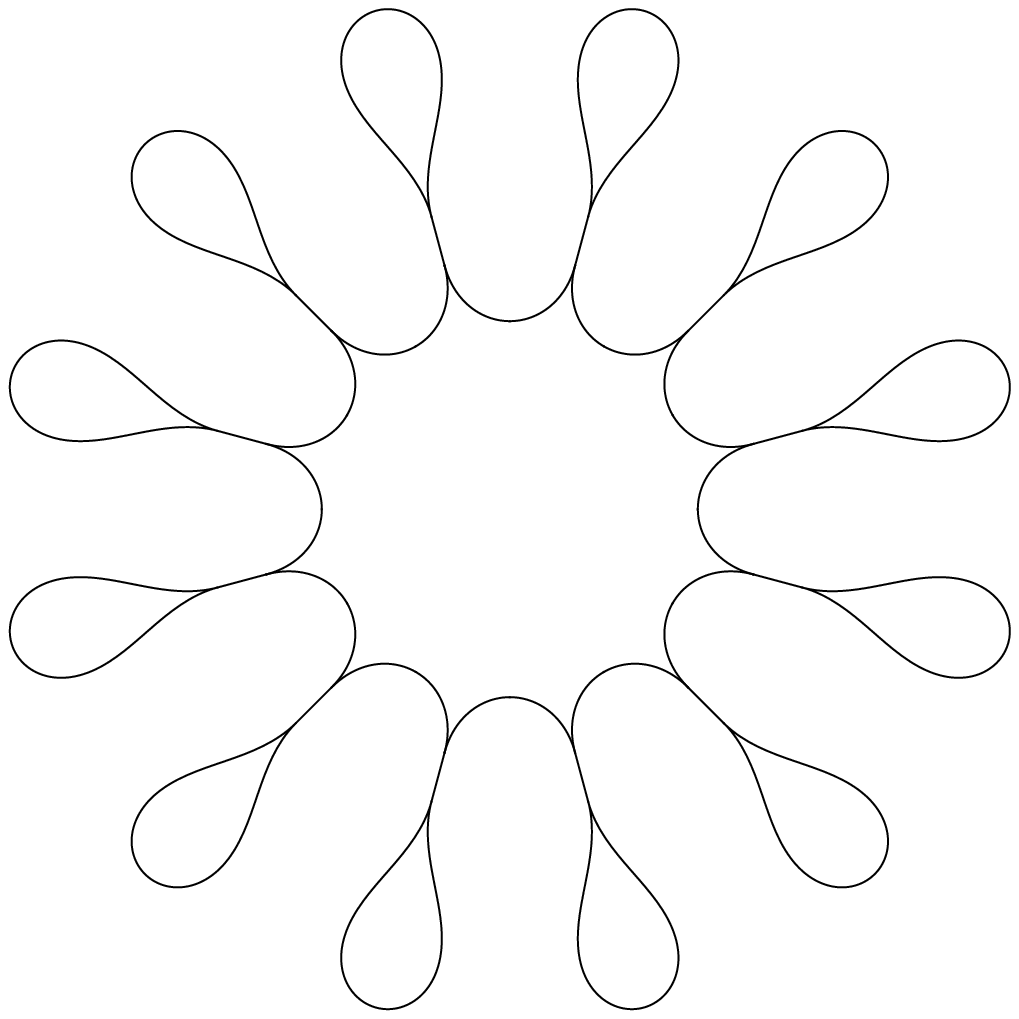}
\caption{Shapes with lines of contact corresponding to:
(a) $\sigma = 400$ (6-fold symmetry); (b) $\sigma = 800$ (9-fold symmetry); (c) $\sigma = 1500$ (12-fold symmetry).}
\end{figure}

It is clear that the tangent, normal and position vectors of a shape $\hat{\Gamma}$ with lines of contact constructed in the foregoing way are continuous at each point of the curve $\hat{\Gamma}$. However, its curvature suffers jumps at the end points of the line segments used to substitute the contact points of the respective auxiliary curve $\hat{\Gamma}_{cn}$ because the limit values of the curvature from the bent parts of the curve and from the line segments are $-\sqrt{2\mu}\neq0$ and zero, respectively.
Consequently, the moment and force also suffer jumps at the aforementioned points since their limit values from the bent parts of the curve are
\begin{equation}
M_b=-D \left(\sqrt{2\mu}+\kappa^{\circ} \right), \quad \,\,
N_b=0, \quad \,\,
Q_b=\pm D \sqrt{P\left(-\sqrt{2\mu}\right)},
\label{BMF}
\end{equation}
while along each line of contact the resultant pressure is zero and
\begin{equation}
M_l=-D\kappa^{\circ}, \qquad
N_l=0, \qquad
Q_l=0.
\label{LMF}
\end{equation}
Equations (\ref{BMF}) and (\ref{LMF}) follow by the constitutive equation (\ref{eq14}), the general solution (\ref{NQ}) of Eqs.~(\ref{Eqil1}) -- (\ref{Eqil3}) and Eqn.~(\ref{JPhysA}).

Thus, the local balances (\ref{eq11}) and (\ref{eq13}) of the force and moment are violated for the shapes with lines of contact. Fortunately, however, the total balances
\begin{equation}
\oint_{\hat{\Gamma}} \mathbf{F}'(s) \mathrm{d}s =
-\oint_{\hat{\Gamma}} p\, \mathbf{n}(s) \mathrm{d}s,
\label{eq11i}
\end{equation}
\begin{equation}
\oint_{\hat{\Gamma}} M'(s) \mathrm{d}s = -\oint_{\hat{\Gamma}}\mathbf{F}(s)\cdot \mathbf{n}(s) \mathrm{d}s,
\label{eq13i}
\end{equation}
of these quantities are satisfied. Indeed, Eqs.~(\ref{eq11i}) and (\ref{eq13i}) hold on the curve $\hat{\Gamma}_{cn}$ since it corresponds to an equilibrium shape without jump discontinuities of the force and moment. On the other hand, $\hat{\Gamma}=\hat{\Gamma}_{cn}$ $\cup$ \{Line segments\} and the integrals in Eqs.~(\ref{eq11i}) and (\ref{eq13i}) taken along the line segments are equal to zero because here $p=0$, $M'(s)=0$ and $\mathbf{F}(s)=\mathbf{0}$.

In our opinion, this property of the constructed curves $\hat{\Gamma}$ allows this shapes to be regarded as equilibrium ring (tube) shapes with lines (areas) of contact at least in the week sense discussed above.

In the light of the results presented in this Section, it should be remarked that the similarity law (5.4) obtained in \cite[Section~5]{Flaherty}, which concerns the conductivity of a buckled tube  conveying an incompressible viscous fluid, has to be revised. Actually, this law, which expresses the conductivity of a tube with areas of contact through that of a tube whose cross sections have just points of contact, should be replaced by the following one
\begin{equation}
\mathcal{C}(\sigma)=\left( \frac{\sigma_{cn}}{\sigma} \right)^{4/3} \mathcal{C}(\sigma_{cn}), \qquad \sigma>\sigma_{cn},
\label{SML}
\end{equation}
where $\mathcal{C}(\sigma)$ denotes the conductivity of a tube subject to pressure $\sigma$. It is necessary to do so because $\sigma_{cn}$ is the unique pressure for which there exists an equilibrium tube shape of $n$-fold symmetry whose cross sections have only isolated points of contact.

\section{Concluding remarks}

In the present paper, the problem for determination of the equilibrium shapes of a circular inextensible elastic ring (tube) subject to a uniform hydrostatic pressure is reexamined.
For the first time, more than a century ago, this problem was stated and studied by Maurice L\'{e}vy in his memoir \cite{Levy}.
 
Here, a concise derivation of the most important facts established in \cite{Levy}, see $1^{\circ }$ -- $3^{\circ }$ (p. 9), concerning the existence of a ``centre of the elastic forces", following by Eq.~(\ref{Fxy}), and the properties reflected by Eqs.~(\ref{Fpr}) and (\ref{BMC}) is given in Section 3.
Then, the parametric equations of the equilibrium shapes are expressed through the forces and slope angle, see Eqs.~(\ref{parametric}). 

It is noteworthy that neither the relations (\ref{Fxy}) -- (\ref{BMC})
nor the forms of the parametric equations (\ref{parametric}) or boundary conditions (\ref{BC1}) and (\ref{BC2}) depend on the particular material properties or stress-free configuration of the ring.

A concise justification of the symmetry of the ring shapes discussed by many authors (usually without going into much detail) is given in Section~4. 
It is shown analytically that each ring shape is symmetric and can be obtained by successive reflections of its part corresponding to the first half period of the curvature with respect to its symmetry axes.

Further, assuming that the stress-free configuration of the ring is a circle of radius $\rho$, the case in which the linear constitutive equation (\ref{eq14}) holds is considered. In this case, the equilibrium state of the ring is determined by the periodic solutions of the nonlinear ordinary differential equation (\ref{JPhysA}) for the ring curvature, which are such that the closure condition (\ref{closure}) holds.
In fact, the shape of the ring is determined explicitly by the parametric equations (\ref{parametric1}) and the values of the moment and forces acting along the ring axes are given by Eqs.~(\ref{eq14}) and (\ref{NQ}), respectively.
Explicit analytic expressions for all periodic solutions of Eq.~(\ref{JPhysA}) and for the corresponding slope angles
are presented in Section~5.

In Section~6, the determination of the equilibrium shapes corresponding to a given pressure $\sigma$ is reduced to the computation of the common solutions of two transcendental equations (\ref{closure1}) and (\ref{L}).

Shapes with isolated points of contact are studied in Section~7. It is shown that the pressures at which such a shape is attained can be obtained computing the common solutions of Eqs.~(\ref{closure1}), (\ref{L}) and (\ref{sigma_c}) in the case $n=2$, or Eqs.~(\ref{closure1}), (\ref{L}) and (\ref{cs1-}) for $n>2$.

The most important results achieved in Sections 6 and 7 are as follows.
In contrast to the assertion in \cite{Flaherty} that for each mode $n$ there exists a range of pressures for which the respective ring shape has only isolated points of contact,
we found, solving numerically Eqs.~(\ref{closure1}), (\ref{L}) and (\ref{sigma_c}) or (\ref{cs1-}), that for each mode $2 \leq n \leq 15$ there is a unique such pressure, namely $\sigma_{cn}$.
Moreover, for all modes in the range $2 \leq n \leq 15$ our computations based on the procedure described in Section~6 show that if the applied pressure $\sigma$ is such that $\sigma_{bn}<\sigma<\sigma_{cn}$, then the corresponding buckled shape of $n$ mode is simple, while for $\sigma>\sigma_{cn}$ this shape always has points of self-intersection.
Our conjecture is that this behaviour is inherent to all modes.

Section 8 concerns the equilibrium ring (tube) shapes with lines (areas) of contact that are expected to occur for pressures greater than the respective contact pressure instead of the self-intersecting shapes (unnatural for tubes and planar rings) predicted by the considered model. Here, the construction of these shapes is based on the similarity properties of Eqs.~(\ref{JPhysA}) and (\ref{parametric1}) following in general outline the idea suggested by Flaherty {\it et al.} in \cite{Flaherty}, but the uniqueness of the contact pressures $\sigma_{cn}$ is taken into account. The shapes obtained in this way are shown to satisfy the total balances (\ref{eq11i}) and (\ref{eq13i}) of the respective forces and moments, which is good reason to consider them as equilibrium shapes. Finally, the expression for the so-called similarity law obtained in \cite{Flaherty} is revised, see Eq.~(\ref{SML}).

The interested reader can find the \texttt{Mathematica}$^{\circledR}$ notebooks developed for the solution of the transcendental equations of Sections 6 and 7 as well as the notebooks developed for the construction of the shapes described in Section 8 at
\texttt{http://www.bio21.bas.bg/ibf/dpb\_files/mfiles/}.

\section*{Acknowledgements}

This research is supported by the contract \# 35/2009 between the Bulgarian and Polish Academies of Sciences.

\end{document}